\documentclass{aa}  

\usepackage{graphicx}
\usepackage{txfonts}
\usepackage{xcolor}
\usepackage[]{hyperref}
\newcommand{\cmnt}[1]{}

\newcommand{\porb}{$P_\mathrm{orb}$}

\newcommand{\mdot}{$\dot{M}$}

\newcommand{\numconf}{123}
\newcommand{\numcand}{29}
\newcommand{\numamcvn}{96} 
\newcommand{\numhecv}{15} 
\newcommand{\numporb}{101} 
\newcommand{\numgaia}{111} 
\newcommand{\numgaiacand}{19}

\begin{document}

   \title{A Catalogue of Ultracompact Mass-Transferring White Dwarf Binaries}


   \author{Matthew J. Green\inst{1,2,3}
          \and
          Jan van Roestel\inst{4}
          \and
          Tin Long Sunny Wong\inst{5}
          }

   \institute{Max-Planck-Institut f\"{u}r Astronomie, K\"{o}nigstuhl 17, D-69117 Heidelberg, Germany\\
              \email{mjgreenastro@gmail.com}
         \and
             Homer L. Dodge Department of Physics and Astronomy, University of Oklahoma, 440 W. Brooks Street, Norman, OK 73019, USA
        \and
             JILA, University of Colorado and National Institute of Standards and Technology, 440 UCB, Boulder, CO 80309-0440, USA
         \and
             Anton Pannekoek Institute for Astronomy, University of Amsterdam, 1090 GE Amsterdam, The Netherlands\\
             \email{j.c.j.vanroestel@uva.nl}
         \and
             Department of Physics, University of California, Santa Barbara, CA 93106, USA\\
             \email{tinlongsunny@ucsb.edu}
             }

   \date{Received ??; accepted ??}

 
  \abstract{
We present an overview and catalogue of all known mass-transferring binary systems that have white dwarf accretors and orbital periods shorter than 70 minutes, including the AM\,CVn-type binary systems and a number of related objects.
In this paper, we provide an overview of the literature background and describe recent breakthroughs, open questions, and connections to other fields.
This overview is intended to provide a starting point for researchers who are relatively new to the fields of ultracompact binaries or AM\,CVn-type binaries.
We define the scope of the catalogue (the criteria for inclusion) and describe its format and contents (what data have been included for each target).
We also summarise the observed properties of the known mass-transferring ultracompact binaries, including their orbital periods and component stellar masses and radii.
The catalogue presented here aims to provide a central resource to enable researchers to track the current status of the known sample.
At the time of writing, the catalogue includes \numconf\ confirmed binaries and \numcand\ candidates.
The catalogue is hosted on a public platform (Zenodo) with version control, and will be updated regularly.
  }

   \keywords{binaries: close; stars: dwarf novae; novae, cataclysmic variables; white dwarfs; catalogs}

   \maketitle
%

\section{Introduction}

65 years ago, \citet{Greenstein1957} took a spectrum of what they believed to be a peculiar, helium-atmosphere white dwarf\footnote{Having been identified as a ``decidedly blue'' star by \citet{Humason1947}, leading to the star's earlier name of HZ\,29.}.
The spectrum showed a series of broad helium absorption lines, with apparently double cores.
Several years later, \citet{Smak1967} observed a photometric signal with a period of 18\,min. 
He suggested the interpretation that the object is an unusually compact binary system, with the photometric period corresponding to the orbit of the binary.
The object was given a variable star name AM Canum Venaticorum (AM\,CVn)\footnote{Over five decades later, the same \citet{Smak2023} combined his original 1962 data with more recent observations to measure the period derivative of the binary system.}.
Over the next two decades, variable objects GP\,Com \citep{Warner1972} and CR\,Boo \citep{Nather1985} were proposed to be binary systems of a similar nature.

Over time, our present understanding emerged: there exists a class of mass-transferring, `ultracompact' binary systems (mass-transferring UCBs), with orbital periods in the range 5--70\,min. 
Each of these binary systems consists of a central white dwarf accreting material via Roche lobe overflow (RLOF) from a low-mass donor, which is either degenerate or semi-degenerate.
The degenerate natures of the donor stars are necessitated by the orbital periods, which are shorter than can be achieved by mass-transferring binaries with non-compact donor stars \citep[e.g.][]{McAllister2019}.
They can be thought of as double-degenerate counterparts to cataclysmic variables \citep[CVs; e.g.][]{Warner1995} or as mass-transferring counterparts to double white dwarf binaries \citep[DWDs; e.g.][]{Maoz2018,Munday2024}.

The accreted material is dominated by helium and often enriched by metals such as nitrogen \citep[e.g.][]{Marsh1991}.
In most cases hydrogen is not detected at all. 
Several of the systems that are often termed AM\,CVn-type binaries, which we take here to mean systems with no detectable hydrogen, have upper limits of order $10^{-4}$ on the hydrogen fraction \citep{Nagel2009,Green2019}. 
However, there are a number of mass-transferring UCBs that do have detectable levels of hydrogen, as discussed later.

\begin{figure}
\centering
\includegraphics[width=\hsize]{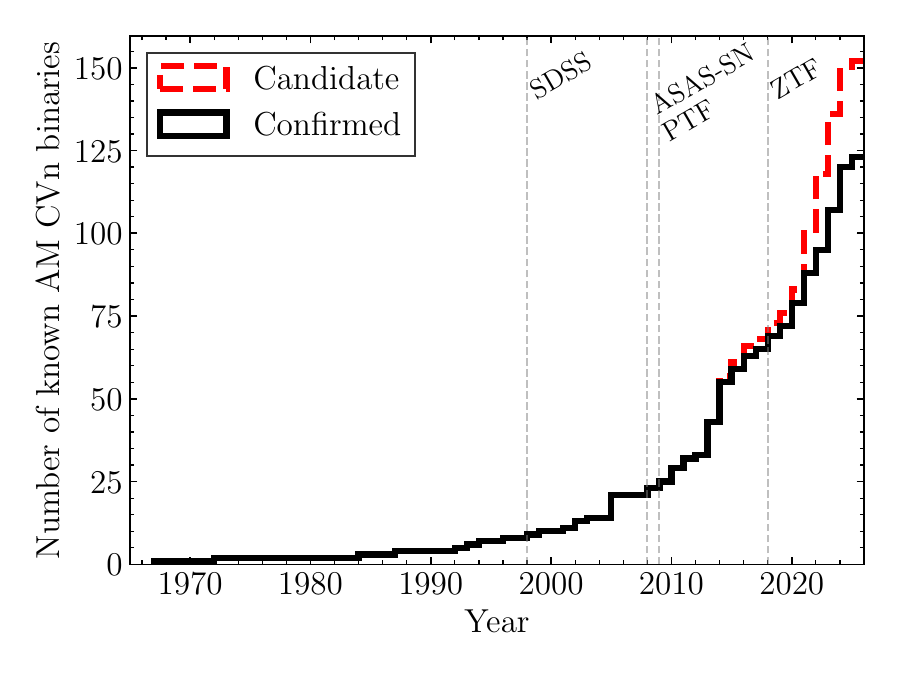}
\caption{The rise in the number of known mass-transferring UCBs since \citet{Smak1967} first suggested that the variable star AM\,CVn may be a 17\,min mass-transferring binary system.
Also plotted are the times at which several large sky surveys went live, contributing to the rapid growth in the known sample since the turn of the century.
}
\label{fig:discovery-year}
\end{figure}

For many years the number of known mass-transferring UCBs grew slowly: until the year 2000, less than ten mass-transferring UCBs were known. 
The dawn of large surveys has accelerated that growth. 
By the review of \citet{SolheimAMCVn}, the number of known systems was 25, and that increased to 56 by the time of the population study by \citet{Ramsay2018}.
The present catalogue contains \numconf\ confirmed systems and a further \numcand\ candidates. Fig.~\ref{fig:discovery-year} shows the growth of the known population over time, and the influence of several large-area surveys.

For the first time, the known population is large enough to derive population-level statistics, a promising avenue for testing formation models.
This will require targeted follow-up of the known systems, especially those from which the most information can be learned (such as bright or eclipsing systems).
Given the accelerating growth of the known population, it is becoming harder for individual researchers to keep track of all known systems.
We therefore present a catalogue of all known mass-transferring UCBs, with the intention to update the contents at regular intervals.
We hope that researchers will use this catalogue as a starting point for observing campaigns or as a basis for population-level statistical studies.

In the following section we give a brief review of our current understanding of mass-transferring UCBs.
The catalogue itself and an overview of the systems that it contains are described in subsequent sections.

Throughout this work we will use the term `UCB' to refer to any binary system with an orbital period $< 70$\,min; the terms `AM\,CVn-type binary' or `classical AM\,CVn binary' to refer specifically to any system with no hydrogen visible in its optical spectrum; and `helium cataclysmic variable' (He~CV) to refer to any mass-transferring UCB that is significantly helium enriched, but still has clear hydrogen lines in its optical spectrum.
The need to distinguish between the latter two sub-classes is driven by a number of observational differences, which we discuss in Section~\ref{sec:diversity}.
The first-discovered binary, after which the class is named, will be referred to as `AM\,CVn itself', to avoid confusion with the class name.

\section{Background and overview}
\label{sec:overview}

\subsection{Accretion behaviour}
\label{sec:accretion}

\begin{figure}
\centering
\includegraphics[width=\hsize]{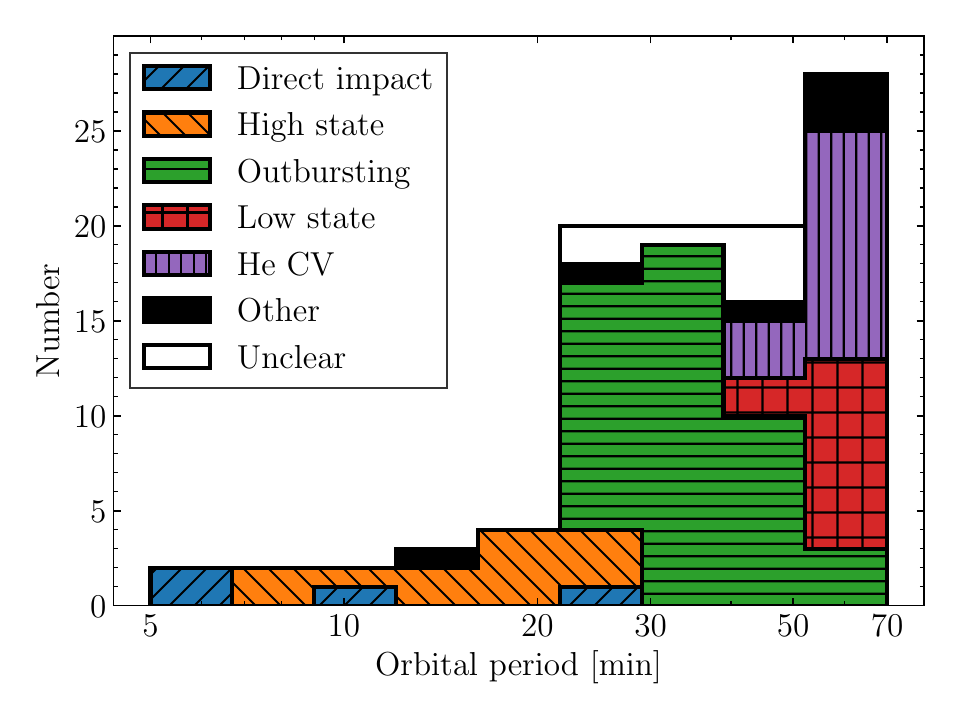}
\caption{The period distribution of all known mass-transferring UCBs, coloured according to their accretion disc behaviour in the case of AM\,CVn-type systems (those without detected hydrogen), with He~CVs (systems with detected hydrogen) and other classes coloured separately.
The accretion disc behaviour clusters by orbital period, but with some regions of overlap between different sub-classes.
He~CVs are typically found at periods $\gtrsim 50$\,min.
Note that most He~CVs do show outbursts, while systems in the same period range without hydrogen typically do not.
The orbital period distribution is strongly influenced by selection effects, most notably the fact that it is easier to find outbursting systems.
}
\label{fig:periods}
\end{figure}

Most mass-transferring UCBs have an accretion disc centred on the accreting white dwarf. 
The physics of a helium-dominated accretion disc has been explored by several works \citep[e.g.][]{Kotko2012,Cannizzo2015,Coleman2018,Cannizzo2019}.
The observed population can be broken into several categories by the behaviour of the accretion disc: whether it is in a permanently bright and ionised `high state', a permanently cool and faint `low state', or alternates between the two through a cycle of dwarf nova outbursts\footnote{More properly, most mass-transferring UCBs experience `super-outbursts' more often than regular outbursts, though both of these phenomena as well as a variety of other behaviours are observed \citep[e.g.][]{Duffy2021,Marcano2021}. Superoutbursts are somewhat brighter and longer in duration than regular outbursts. For the purposes of this work we will not emphasize the distinction.} 
These categories are grouped relatively clearly by orbital period (Fig.~\ref{fig:periods}).

These trends can be explained by considering the large dynamical range of allowed mass transfer rates, and the steep dependence of mass transfer rate on orbital period.
The dominant form of angular momentum loss is believed to be gravitational wave radiation, which depends strongly on orbital period: under the assumption of a zero-temperature white dwarf donor, it can be shown that the mass transfer rate depends on orbital period as $\Dot{M} \propto P_{\rm orb}^{-14/3}$ \citep[for a more in-depth discussion, see][]{Cannizzo2015}.
At short periods ($P_{\rm orb} \lesssim 20$\,min), the high $\dot{M}$ leads to a hot, stable, and ionised accretion disc. 
As $\dot{M}$ decreases with increasing $P_{\rm orb}$, the disc encounters a thermal-viscous instability and exhibits dwarf nova outburst behaviour.
Among systems that show dwarf nova outbursts, the outburst recurrence rates, durations, and amplitudes show strong dependences on orbital period \citep{Levitan2015,Cannizzo2015,Cannizzo2019}.
At even longer orbital periods ($P_{\rm orb} \gtrsim 50 - 60$\,min), $\dot{M}$ is too low to drive dwarf nova outbursts, and the disc exists in a persistent low-state. 
The boundaries of the classes in period space are somewhat fuzzy, which may be either evidence for a diversity of masses of the accretor stars, evidence for a range of levels of entropy in the donor stars, or evidence that instantaneous \mdot\ values can deviate somewhat from the expected secular \mdot.

During outbursts, a system brightens by 2--6 magnitudes for a duration of 10--100 days.
Very often a periodic variation is seen during outburst, termed the `superhump' period \citep[e.g.][]{Patterson1993,Kato2013}. 
This originates from an interaction between the orbit and the precessing disc (the exact nature of the interaction is debated), and occurs at the beat frequency between the disc precession and orbital frequencies.
As this is typically within a few per cent of the orbital period, it can be used as a proxy for systems for which the orbital period has not been measured directly.

The superhump frequency may allow some insight into the nature of the system.
For unevolved, hydrogen-transferring CVs, an empirical relationship has been found between the superhump period, orbital period, and the mass ratio.
Several forms of this relationship have been put forward \citep[e.g.][]{Patterson2005,Knigge2006,Kato2013,McAllister2019}.
It is common practise to apply these same relations to helium-transferring UCBs in order to derive the mass ratio, but caution should be used: these relations are not necessarily well calibrated for discs with these compositions \citep[e.g.][]{Pearson2007}.
For YZ\,LMi the superhump-derived $q$ agrees well with the eclipse-derived value \citep{Copperwheat2011a}, while for AM\,CVn itself, the spectroscopic derivation disagrees with the superhump-derived value \citep{Roelofs2006b}.
Further investigation of systems that show both eclipses and outbursts will provide further insight.

A separate class of brightening event is also observed among AM\,CVn-type binaries, sometimes termed `long-duration outbursts' or `long-period outbursts' \citep{RiveraSandoval2020,RiveraSandoval2021,RiveraSandoval2022,Wong2021b}. 
These brightening events have to date only been observed among long-period AM\,CVn-type binaries ($P_{\rm orb} \gtrsim 50$\,min).
During these events the system brightens by 1--3 mags for a duration of 100-400 days.
Although they are termed outbursts, the observed properties of these events are substantially different to the more typical dwarf nova outbursts: they are longer and fainter, and are characterised by colours that are redder than the quiescent colour of the binary (while dwarf nova outbursts are typically bluer).
The binary ASASSN-21au has been observed to undergo both long-duration outbursts and dwarf nova outbursts \citep{RiveraSandoval2022}.
The physical nature of these long-duration outbursts is not yet well understood, but it has been suggested that a temporary enhancement of mass transfer may be necessary, as the duration of these events exceeds the viscous timescale of the accretion disc \citep{RiveraSandoval2020}.

At the  shortest orbital periods (typically $P_{\rm orb} \lesssim 10$\,min), the binary is too compact for the accretion disc to form, and the accreted material impacts directly on the white dwarf.
The prototypical examples of these systems are HM\,Cnc and V407\,Vul \citep{Motch1996,Israel1999,Ramsay2002}.
While several models were proposed to explain the nature of these systems \citep{Motch1995,Cropper1998,Israel1999,Wu2002}, the direct-impact UCB model was best able to explain the photometric and spectroscopic observations of the systems \citep{Barros2007,Roelofs2010} and had become broadly accepted by the time of the review by \citet{SolheimAMCVn}.
The spin of the accretor can act as an orbital angular momentum sink removing angular momentum from the orbit, but the stability of these systems is made possible by a tidal torque between the accretor and donor stars \citep{Marsh2004,Gokhale2007,Kremer2015}.
The orbital period boundary between UCBs with and without accretion discs is not strict: \citet{Chakraborty2024} found two disc systems with $P_{\rm orb} \approx 8$\,min, making them more compact than the disc-less V407\,Vul ($P_{\rm orb} = 9.5$\,min), while a direct impact system has even been claimed with a period of 23\,min by \citet{Ramsay2018b}

\subsection{Formation scenarios}
\label{sec:evolution}

\begin{figure}
\centering
\includegraphics[width=\hsize]{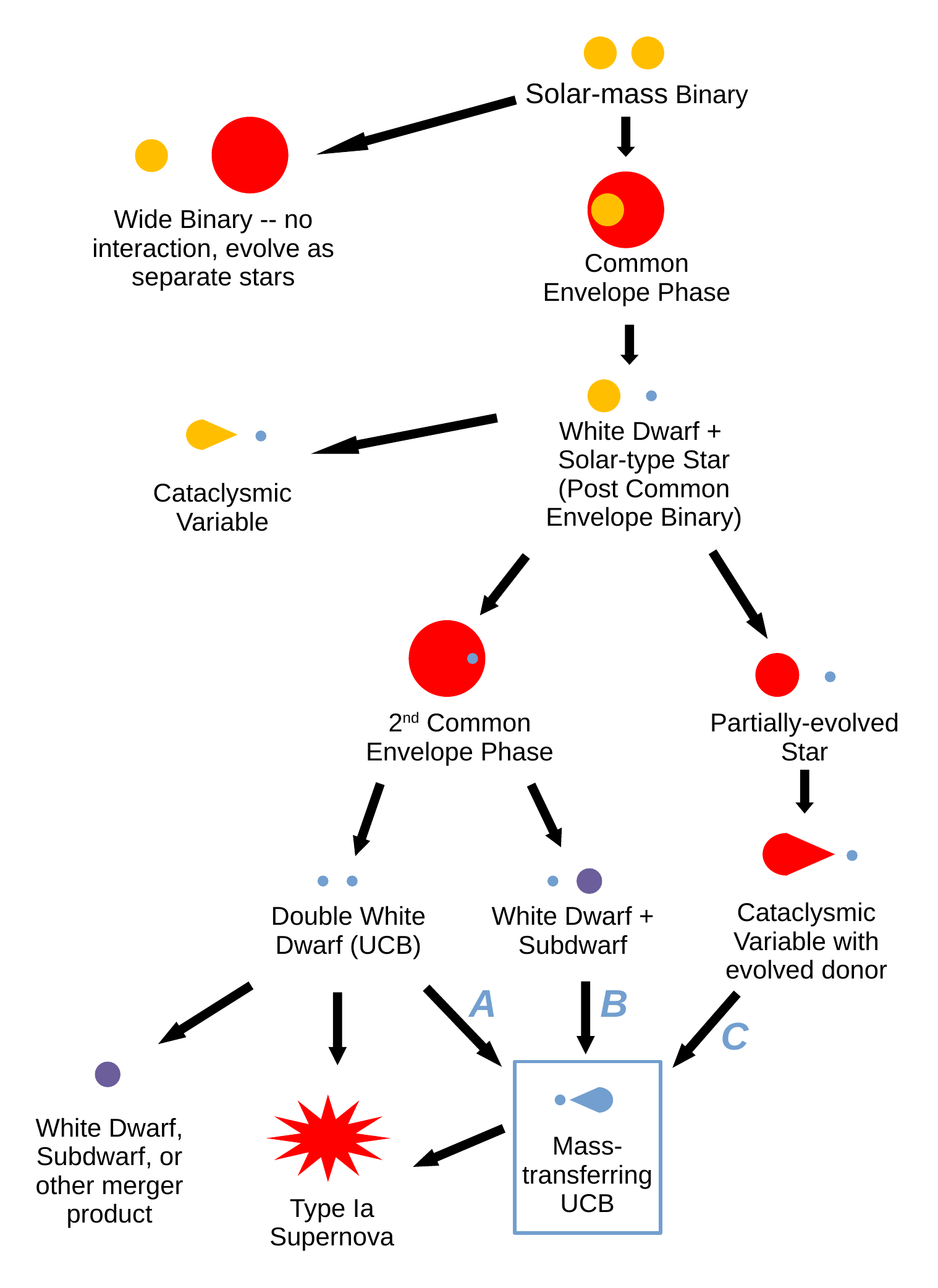}
\caption{A schematic diagram of the proposed evolutionary paths by which mass-transferring UCBs can form, beginning with a Solar-type binary system and passing through three proposed formation channels. Labels highlight the white dwarf donor channel (A), helium star donor channel (B), and evolved cataclysmic variable channel (C). Details of these channels are discussed in the text.
}
\label{fig:evolution}
\end{figure}

One of the prominent open questions in the study of mass-transferring UCBs is the question of their formation. 
Proposed formation models have generally coalesced into three channels: the white dwarf donor channel, the helium star channel, and the evolved CV channel. 
A schematic version of these evolutionary channels is presented in Fig.~\ref{fig:evolution}.

In the white dwarf donor channel \citep{Paczynski1967,Deloye2007,Wong2021,Chen2022}, the binary system passes through two phases of common envelope evolution as both stars evolve off the main sequence, leaving two white dwarfs, one of which is low mass. 
Following a period of inspiral, mass transfer starts at an orbital period of $\approx 2$--10\,min \citep{Wong2021}.
If mass transfer is stable, the period evolution turns around after a short time to evolve outwards, moving quickly through the short orbital period range and more slowly through longer orbital periods.
If mass transfer is unstable, then the binary will merge to form a high-mass white dwarf, hot subdwarf, or R\,CrB star; or may form a Type Ia supernova (Section~\ref{sec:gwsn}).
The peak mass transfer rate (shortly after the onset of mass transfer) is high enough in some systems to trigger helium nova eruptions on the surface of the accretor, which have the potential to destabilise the binary system \citep{Shen2015} or cause a thermonuclear detonation of the accretor \citep{Wong2023}.

The helium star donor channel \citep{Savonije1986,Iben1987,Yungelson2008,Bauer2021,Sarkar2023a,Rajamuthukumar2024} follows a similar path, with the main distinguishing feature being that the donor star is burning helium in its core at the onset of mass transfer.
This results in a more inflated donor star and pulls the onset of mass transfer towards slightly longer orbital periods of $\approx 10$--15\,min.
Both helium novae and Type Ia supernovae are also possible for systems with helium star donors \citep[e.g.][]{Bauer2017,Rajamuthukumar2024}.
It may be possible to distinguish donors formed by this channel from others by their chemical composition, in particular by the ratio of nitrogen to carbon \citep[e.g.][]{Yungelson2008,Nelemans2010}.
Non-mass-transferring WD+sdB binaries with short periods \citep[e.g.][]{Geier2013,Lin2024} may be progenitors to mass-transferring binaries formed through this channel.

The evolved CV channel \citep{Tutukov1985,Podsiadlowski2003,Goliasch2015,Sarkar2023,Belloni2023} is a rather different process: the binary system passes through only one period of common envelope evolution, forming the central white dwarf, and begins stable mass transfer with a donor that is somewhat evolved off the main sequence.
The outer layers of the donor are then stripped by this stable mass transfer, revealing the helium core, and the accreted material becomes helium dominated.
Historically, models of this channel struggled to reproduce several features of the system, requiring significant fine-tuning to create binaries with periods $\lesssim 30$\,min or without detectable levels of hydrogen \citep[e.g.][]{Goliasch2015}.
Recent work has shown that the problems of historic models can be resolved by adopting magnetic braking prescriptions with higher rates of angular momentum loss.
The Convection And Rotation Boosted (CARB) magnetic braking prescription, originally developed for X-ray binary evolution \citep{Van2019}, also successfully produces AM\,CVn-type binary systems when applied to cataclysmic variables with evolved donors \citep{Belloni2023}.
Another possibility is the double-dynamo magnetic braking prescription, which increases the strength of the magnetic field on the donor star, and can also produce UCBs \citep{Sarkar2023}.
However, it should be noted that these enhanced models of magnetic braking, in particular the CARB model, do not well describe populations of single stars, or binary systems consisting of two main sequence stars \citep[e.g.][]{El-Badry2022}.

All of these channels may contribute some fraction to the total UCB population, but the fraction of systems forming from each channel remains an area of active research.
Each channel can recreate the parameters of most individual systems, and it is likely that the best test of formation models will come from a comparison of population statistics.

The eventual fate of the mass-transferring UCBs is somewhat uncertain.
In principle, evolutionary models suggest that mass-transferring UCBs can evolve to $P_\mathrm{orb} > 70$\,min within a Hubble time \citep[e.g.][]{Wong2021}.
However, the survey for eclipsing systems by \citet{vanRoestel2022} did not find helium-transferring binaries at periods longer than 70\,min, even though such systems would have been within the survey's sensitivity.
If mass transfer were to cease at a period of $\approx 70$\,min and if the donor were to survive, the next stage of the binary would appear like a white dwarf--brown dwarf binary system in which the white dwarf has a helium atmosphere.
This has also been predicted in the context of ultracompact X-ray binaries (see Section~\ref{sec:ucxbs}) by \citet{Sengar2017}.
However, although several white dwarf--brown dwarf binaries were found at similar periods in the survey presented by \citet{vanRoestel2022}, none contains a white dwarf with a helium atmosphere.

\subsection{Gravitational waves and Type Ia supernovae}
\label{sec:gwsn}

Besides the evolutionary questions discussed above, studies of mass-transferring UCBs are often motivated by either their strong emission of gravitational waves or by their evolutionary connections to Type Ia supernovae.

UCBs (both mass-transferring and non-mass-transferring) are loud sources of gravitational wave radiation in the frequency band that will be visible to space-based interferometers such as the Laser Interferometer Space Antenna \citep[LISA;][]{Amaro-Seoane2017} and TianQin \citep{Luo2016,Ye2019}.
The shortest-period UCBs are expected to be among the loudest Galactic sources visible to LISA \citep{Kremer2017,Breivik2018}, and they are planned to be used as calibration and verification sources \citep{Kupfer2024}.
Approximately half of the proposed calibration sources are mass-transferring binaries.

Longer-period UCBs will contribute to an unresolved gravitational wave foreground which sets the detection limit of LISA at certain frequency ranges. 
In general it is assumed that non-mass-transferring DWDs, which make up the majority of UCBs by space density, will be the dominant source of this unresolved foreground \citep[e.g.][]{Korol2017,Korol2022}.
While the contribution of mass-transferring systems is likely to be weaker, its exact strength is difficult to quantitatively predict given current uncertainties around the population and its evolution.

Double-degenerate binary systems have long been thought of as candidate progenitors of type Ia supernovae, and in the last decade this scenario has gained in popularity.
Reviews of Type~Ia supernova formation and mechanisms are given by \citet{Maoz2014} and \citet{Patat2018}, but see also work on the `dynamically driven double-degenerate double-detonation' scenario by \citet{Shen2018b,Shen2018a}.
In general, a Type~Ia supernova can arise from a double-degenerate binary if the two white dwarfs merge to form a super-Chandrasekhar mass white dwarf, or if a helium detonation on the surface of the accreting white dwarf causes a pressure wave that triggers detonation at its core (the `double-detonation' scenario).
The double-detonation scenario can occur either during dynamically-driven accretion just before a merger \citep[e.g.][]{Guillochon2010,Shen2015,Shen2018b} or during stable mass transfer \citep[in which case the surface detonation arises during a strong helium flash, e.g.][]{Bildsten2007,Wong2023}, but in both cases it is crucial that the accreted material be dominated by helium.
It should be noted that \citet{Piersanti2014,Piersanti2015} have argued that the stable mass transfer case of double-detonation should occur only rarely.
In the double-detonation scenario, the accretor typically needs to be $\approx 1.0 M_\odot$ in order to produce a typical-luminosity supernova.
Masses of $\gtrsim 1.1 M_\odot$ can produce over-luminous supernovae (provided the core is still carbon-oxygen), while masses of $\approx 0.8$--$0.9 M_\odot$ may produce subluminous supernovae \citep[e.g.][]{Shen2018a,Shen2021}.

A minority of systems in this catalogue may themselves be Type~Ia supernova progenitors. 
In particular, \mbox{ATLAS\,J1138$-$5139} \citep{Chickles2024,Kosakowski2024} hosts a $1.0 M_\odot$-mass primary star and is believed to still be moving towards shorter periods and higher mass transfer rates.
This system may be a Type~Ia supernova progenitor in the next few million years if the mass transfer rate becomes sufficiently high.
Two systems with subdwarf donors both have relatively low-mass accretors and may experience sub-luminous supernovae \citep{Bauer2021}\footnote{See also \citet{Deshmukh2024} for a discussion of a comparable WD+sdB binary that is not currently mass-transferring, and \citet{Rajamuthukumar2024} for a detailed discussion of novae and supernovae in WD+sdB binaries}.
A greater number of candidate Type~Ia supernova progenitors have been identified among non-mass-transferring DWDs \citep[e.g.][]{Munday2024}.

The majority of the mass-transferring UCBs in this catalogue can be understood as systems in which a Type~Ia supernova did not and will not occur. 
The mass transfer rates in most systems with $P_{\rm orb} \gtrsim 10$\,min are too low to trigger a helium nova eruption.
Additionally, the accretors in most systems with mass measurements are typically $\approx 0.8 M_\odot$, which is too low to produce a typical-luminosity Type Ia supernova.
Studying these systems can still provide some insight into the supernova question: by constraining the initial parameter space of the systems that exist today, we can demonstrate regions of parameter space in which a DWD does not host a supernova.

\subsection{The diversity of mass-transferring UCBs}
\label{sec:diversity}

\begin{figure}
\centering
\includegraphics[width=\hsize]{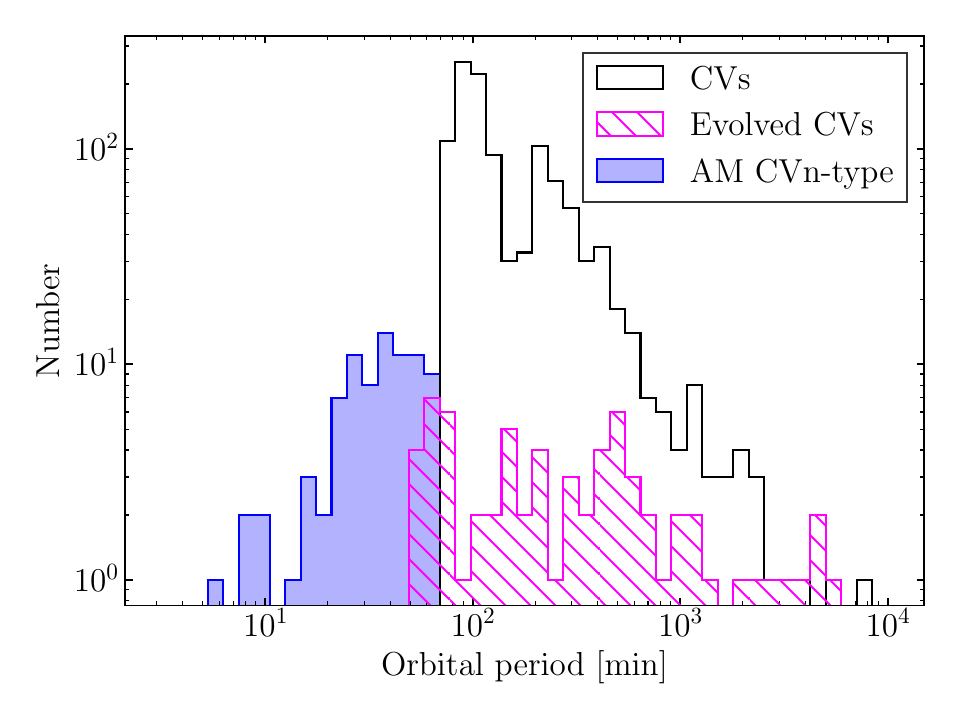}
\caption{The period distributions of the overall population of CVs \citep[][seventh edition, updated in 2014]{Ritter2003}, CVs with evolved donors (provided by B. G\"{a}nsicke, priv.\ comm. 2025), and AM\,CVn-type binaries (systems without detectable hydrogen, this work). Magnetic CVs and CVs with orbital periods below the non-degenerate period minimum were excluded from the CV period distribution.
The orbital periods of evolved CVs cover a broad range, of which the shortest are $\lesssim 70$\,min and therefore fall into the period range of this catalogue.
}
\label{fig:big-period-plot}
\end{figure}


The descriptions given so far apply mostly to what we might term a `classical' AM\,CVn-type system.
Systems of this sub-class make up the majority of accreting white dwarfs with \porb\,$\lesssim 70$\,min, and have a number of observational properties in common with each other.
In addition to the absence of spectroscopic hydrogen lines (which we take as the definition of this subclass), they follow the pattern of accretion disc behaviour with orbital period described in Section~\ref{sec:accretion}, and they form a sequence in terms of their observed donor masses and radii (see Section~\ref{sec:masses}).
However, in the last few decades it has become clear that an impressive diversity of binary systems exist in this period range, including many systems which do not follow these patterns.
This diversity is primarily driven by differences among the donor stars of the binary systems.

Between the classical AM\,CVn binaries and non-evolved CVs are systems that are sometimes called `He CVs', or `hybrid' or `transitional' systems: they have orbital periods of 50--70\,min and contain detectable levels of hydrogen, while still being significantly helium-enriched when compared to typical CVs.
The first-discovered of these was V485\,Cen \citep{Augusteijn1993}, and around 20 of these systems are known today.

These He~CVs are the short-period end of a broad distribution of CVs with evolved donors, which span a much wider range of periods than CVs with unevolved donors (Fig.~\ref{fig:big-period-plot}).
They are naturally connected to the evolved CV formation channel discussed in Section~\ref{sec:evolution}, which supposes that these systems are evolving inwards towards a period bounce, and that at least some fraction will then evolve into the completely hydrogen-depleted AM\,CVn binaries.
The He~CV with the most precise measurements of its component stellar masses and radii is the eclipsing binary ZTF\,J1813+4251, for which MESA evolutionary models do sugget that it will evolve into an AM\,CVn-type system after passing through a period minimum of 20\,min \citep{Burdge2022}.

Two binary systems have been discovered in which the donor is a hot subdwarf \citep{Kupfer2020a,Kupfer2020b}.
These binary systems may also evolve into AM\,CVn binary systems, or they may be potential progenitors of Type~Ia supernovae \citep{Bauer2021,Rajamuthukumar2024}.

Two further unusual systems each consist of a warm donor with some detectable hydrogen in a 14\,min and 28\,min orbit, respectively, around a white dwarf \citep{Burdge2023,Chickles2024,Kosakowski2024}.
The white dwarf donor channel is the favoured formation scenario for both of these systems, and they may either evolve into AM\,CVn-type systems or detonate as Type~Ia supernovae.

Lastly, the catalogue also includes two binary systems with brown dwarf donors.
Both binaries have orbital periods of 66--70\,min, well above the theoretical lower limit of 40\,min predicted by \citet{Nelson2018} for mass-transferring brown dwarf donors.
For the first-discovered of these, SDSS\,J1507+5230, there has been some debate whether the donor descended directly from a brown dwarf \citep{Littlefair2007} or from a metal-poor halo star in which the low metallicity causes the donor to deviate from a typical CV mass-radius track \citep{Szkody2005}.
Measurements of composition and space velocity support the latter scenario \citep{Uthas2011}.
In both scenarios, the donor today could be classed as a brown dwarf.
The recently discovered WD\,J1540$-$3929 also has a brown dwarf donor and may have a similar history \citep{Steen2024}.

There is a remarkable absence of magnetic fields among UCBs.
Among the whole population, only two binaries have been shown to host magnetic fields \citep{Maccarone2024}.
The proposed field strengths in both cases are $\lesssim 100$\,kG, and both show clear observational evidence of an accretion disc.
For comparison, some 30 per cent of hydrogen-dominated CVs in a volume-limited sample \citep{Pala2020} are either polars or intermediate polars, with magnetic fields of order 10s--100s\,MG, which in the case of polars is strong enough to completely disrupt the accretion disc.
The absence of comparable magnetic fields among UCBs may be related to the stability: the stability of accretion during the direct impact phase ($P_{\rm orb} \lesssim 10$\,min) relies on a balance of torques between the accretor and donor \citep{Marsh2004}, and a strong magnetic field may disrupt this balance.
The origin of magnetic fields in CVs is itself an area of active research \citep[e.g.][]{Schreiber2021}, and the inter-relation between magnetic field formation and UCB formation is an aspect that should be considered by researchers in both areas.

\subsection{Ultracompact X-ray binaries}
\label{sec:ucxbs}

This catalogue includes only binary systems in which the accreting object is a white dwarf, but another class of mass-transferring UCB exists: ultracompact X-ray binaries (UCXBs).
\citet{ArmasPadilla2023} give a recent review and a catalogue\footnote{\url{https://www.sc.eso.org/~jcorral/UltraCompCAT/}} of known UCXBs.
In these binaries the accreting object is a neutron star\footnote{A black hole may also be possible, but none has been found so far.}, while the donor is a white dwarf or other (semi-)degenerate object.
They typically have orbital periods $\lesssim 80$\,min, but can have periods as long as 120\,min.
The space density of these objects is much lower than that of white dwarf-containing UCBs, but their strong X-ray emission allows them to be identified at greater distances (2--20\,kpc).
The catalogue of \citet{ArmasPadilla2023} contains 49 systems, of which 20 are confirmed, 25 are candidates, and 4 are related, short-period X-ray binaries.

A number of similarities can be drawn between UCXBs and UCBs containing white dwarfs.
There are three proposed formation channels for UCXBs \citep[e.g.][]{Yungelson2002,Podsiadlowski2002,Nelson2003,VanHaaften2012}, each of which is analaguous to one of the three formation channels proposed for AM\,CVn-type systems (Section~\ref{sec:evolution}).
The donors in UCXBs may therefore be similar to the donors of AM\,CVn-type binaries, but a carbon-oxygen core donor star is also possible and has been proposed for some systems \citep[e.g.][]{Nelemans2010a}.

UCXB formation is significantly enhanced in globular clusters due to dynamical interactions, and 11 of the 49 known UCXBs are members of globular clusters.
Those systems in clusters also have notably shorter orbital periods than the field population.
Two candidate AM\,CVn-type binaries have been found in globular clusters \citep{Zurek2016,RiveraSandoval2018}, but the sample numbers are too small for a direct comparison at present.

\subsection{Large-area surveys and selection effects}
\label{sec:surveys}

The sample of known, mass-transferring UCBs is substantially inhomogeneous: systems in the catalogue have been discovered through a variety of means, in some cases serendipitously and in other cases as a result of targeted searches.
The rapid increase in discovery rate through the last few decades is largely due to targeted searches through data from large-area surveys.
We mark the first-light dates of several of these surveys in Fig.~\ref{fig:discovery-year} to demonstrate this point.
As a result of this history, the known sample of UCBs is subject to a variety of complicated selection effects, which should be considered by any study that utilises the whole sample.

The Sloan Digital Sky Survey \citep[SDSS;][]{York2000,Abazajian2009,Eisenstein2011} has published both spectroscopy and (for a larger number of targets) five-colour photometry.
A number of targets have been directly discovered through their SDSS spectra \citep[e.g.][]{Anderson2005,Roelofs2005,Inight2023a,Inight2023b}.
Additionally, \citet{Roelofs2007b,Roelofs2009}, \citet{Rau2010} and \citet{Carter2013,Carter2013a,Carter2014b,Carter2014a} implemented a spectroscopic follow-up campaign of candidate mass-transferring UCBs that were selected on the basis of their SDSS colours.
That survey was believed to be complete (within the SDSS footprint and their colour selection criteria) to a magnitude of $g < 19$, and 70 per cent complete to $g < 20.5$.

Many mass-transferring UCBs have been discovered due to their dwarf nova outbursts (Section~\ref{sec:accretion}).
Time-domain photometric surveys are ideal for this, such as the Palomar Transient Factory \citep[PTF;][]{Rau2009}, the Catalina Real-time Transient Survey \citep[CRTS;][]{Drake2009}, the All-Sky Automated Search for SuperNovae \citep[ASAS-SN;][]{Shappee2013}, the Zwicky Transient Facility \citep[ZTF;][]{Bellm2019}, and \textit{Gaia} Science Alerts \citep{Hodgkin2021}.
Outbursting UCBs have been discovered through data from all of these surveys \citep[e.g.][]{Levitan2011,Levitan2013,Levitan2014,Breedt2012,Breedt2014,Campbell2015,Kato2015,Kato2021,vanRoestel2021}.
In particular, \citet{vanRoestel2021} implemented a selection based on the colour of the target during outburst which resulted in a relatively high UCB true-positive rate of 25 per cent.
\citet{Kato2021} put forward a method to find UCB candidates from among outbursting binaries on the basis of their outburst lightcurves, which has also produced a large number of candidates.
While the detection of outbursting systems is perhaps the easiest method to find mass-transferring UCBs, it introduces a complicated set of selection effects into the known sample.
Naturally, only systems that show outbursts can be found in this manner (among the classical AM\,CVn-type systems this means orbital periods of 20--50\,min, Fig.~\ref{fig:periods}), and between outbursting systems there is a further bias in favour of systems which outburst more frequently (the shorter period end of the outbursting systems).
On the other hand, longer-period outbursters tend to brighten by a greater degree during outburst \citep{Levitan2015}, and so can be detected at greater distances. 
Overall, these selection effects are difficult to quantify and model.

The high time resolution of ZTF has enabled direct searches for eclipsing UCBs, dramatically increasing their representation among the known sample \citep{Burdge2020,Burdge2022,Burdge2023,vanRoestel2022,Khalil2024}.
An advantage to this approach is that eclipse probability depends only weakly on orbital period for mass-transferring UCBs, so this approach does not introduce a strong period selection except for the magnitude limit applied.

Lastly, mass-transferring UCBs are X-ray sources, and have been discovered through X-ray data from the R\"{o}ntgensatellit \citep[\textit{ROSAT};][]{Truemper1982} and the extended R\"{O}ntgen Survey with an Imaging Telescope Array \citep[\textit{eROSITA};][]{Predehl2021}.
This approach is particularly sensitive to UCBs in a direct-impact accretion configuration, which are the strongest X-ray sources \citep{Motch1996,Israel1999,Haberl2017,Maitra2024}, but has also yielded UCBs with accretion discs \citep[e.g.][]{Rodriguez2023,Schwope2024a}.
\citet{Schwope2024b} and \citet{Rodriguez2024} gave overviews of the X-ray properties of mass-transferring UCBs in relation to other X-ray sources.

\section{Criteria for inclusion}
\label{sec:criteria}

\begin{figure}
\centering
\includegraphics[width=\hsize]{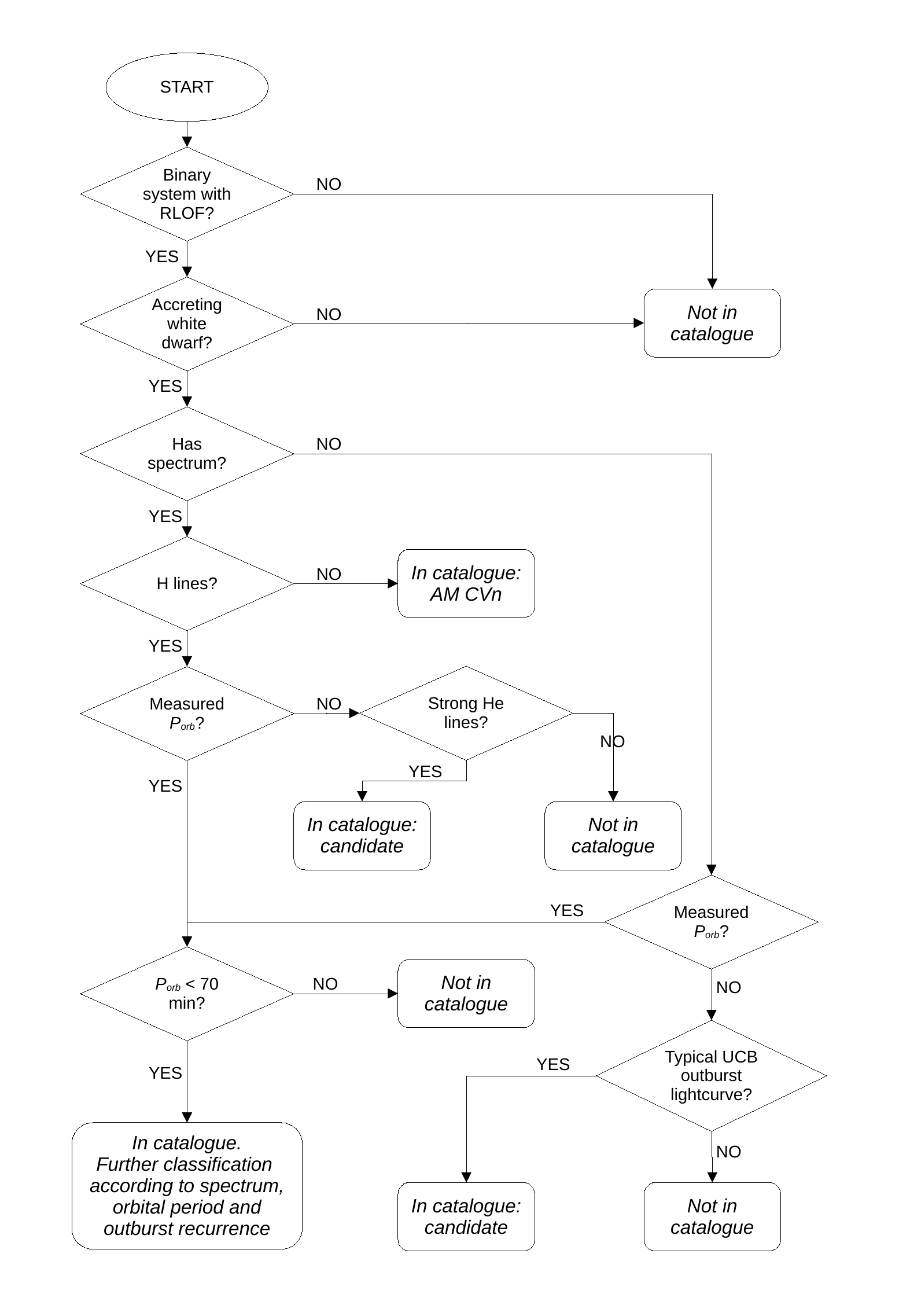}
\caption{Flow chart demonstrating the process by which targets are included in or excluded from the catalogue.
}
\label{fig:flow-chart}
\end{figure}

Several overlapping sets of definitions exist in the literature for the terms `UCB' and `AM\,CVn-type binary'.
These definitions may include a combination of: short orbital periods, the nature of the component stars, the composition of the accreted material, and/or the evolutionary history of the binary system.

We adopt the following set of criteria for inclusion in this catalogue:
\begin{itemize}
    \item The target is a binary system undergoing RLOF.
    \item The accreting object is a white dwarf.
    \item The orbital period of the binary is shorter than 70\,min, or there is reasonable evidence to believe this is the case.
\end{itemize}
Targets that meet these criteria will be referred to as `mass-transferring UCBs' throughout this paper.
These criteria were chosen to maximise the inclusivity of the catalogue while focusing on properties that can be confirmed by observation relatively cheaply.
The existence of mass transfer can be confirmed by the detection of dwarf nova outbursts, spectroscopically-visible disc emission lines, accretion-driven photometric flickering, or (in eclipsing systems) a photometrically-detected bright spot.
The nature of the accreting object as a white dwarf is clear from basic observational properties: mass-transferring binaries such as X-ray binaries appear quite different.

The orbital period (or the superhump period, which serves as a proxy for the orbital period) can often be measured photometrically, especially during an outburst, meaning there is no requirement for more expensive spectroscopic follow-up. 
The orbital period cut-off of 70\,min was chosen so as to include all AM\,CVn-type binary systems \citep[of which the longest orbital period is 68\,min;][]{Green2020}, while avoiding cataclysmic variables with non-degenerate donors \citep[which have a typical period minimum of 80\,min, with scatter down to 76\,min; e.g.][]{McAllister2019}.
As AM\,CVn-type systems have a characteristic spectrum (lacking hydrogen lines) and all are within the period cut chosen here, an AM\,CVn spectroscopic classification is sufficient for inclusion in this catalogue even without a measured orbital period.
The decision process by which targets are included or excluded in the catalogue is illustrated in Fig.~\ref{fig:flow-chart}.

Many discoveries of UCBs are never published in a peer-reviewed journal---early observations are often carried out by amateur astronomers, and discoveries are often only posted to unreviewed forums such as the Variable Stars Network (VSNET)\footnote{\url{http://www.kusastro.kyoto-u.ac.jp/vsnet/}} or Astronomer's Telegram (ATel)\footnote{\url{https://astronomerstelegram.org/}}.
Discoveries may be officially published in peer-reviewed venues only with a significant delay, or never.
Therefore we do include such unreviewed forums as confirmation of the nature of the binary system, if there is (for instance) a clearly reported detection of an outburst with a superhump period $< 70$\,min.

Alongside confirmed systems that meet all of the above criteria, we also include a number of candidate systems. 
The inclusion or exclusion of the candidates is necessarily somewhat more ad-hoc.
Candidates are included when the above criteria have not been confirmed, but there is reason to think they are likely.
Dwarf novae that are suggested to be UCBs due to their outburst shape and duration \citep[the method proposed by][]{Kato2021} are classed as candidates in this catalogue, unless a spectrum or confirmation of the orbital period has been obtained.

\subsection{Sub-class definitions}

We classify the majority of targets in the catalogue into one of several sub-classes of mass-transferring UCBs. 
Here we give the definitions of those sub-classes and the criteria used to identify them.
These sub-classes are discussed in more detail in Section~\ref{sec:diversity}.

\textit{AM\,CVn-type binaries:} We take this to refer to mass-transferring UCBs with no detected hydrogen in their optical spectra.
This includes the majority of confirmed targets in this catalogue (\numamcvn\ out of \numconf).

Several direct-impact systems (HM\,Cnc and the candidate direct-impact system 3XMM\,J0510-6703) do show hydrogen lines.
However, evolutionary theories describing these systems position them as direct progenitors to AM\,CVn-type binaries that are passing through a short-lived phase in which the thin hydrogen shell of the donor is accreted \citep{DAntona2006,Kaplan2012}.
HM\,Cnc also appears to follow the overall trend of AM\,CVn-type donor star masses and radii (Section~\ref{sec:masses}).
Therefore, we consider the direct-impact binaries to belong to the AM\,CVn-type sub-class, despite the presence of hydrogen.

Systems without observed spectra were classed as AM\,CVn-type if they have recorded dwarf nova outbursts and have $P_{\rm orb} < 40$\,min, on the grounds that every UCB exhibiting dwarf nova outbursts in this period range that has been followed up spectroscopically has been found to be an AM\,CVn-type binary.

\textit{He CVs:} We use this term to refer to mass-transferring UCBs that show the presence of hydrogen but still have an enhanced helium abundance relative to the general CV population, and have $P_{\rm orb} < 70$\,min. 
These systems are generally found with $P_{\rm orb} \gtrsim 50$\,min and, as discussed in Section~\ref{sec:masses}, appear substantially different to the AM\,CVn-type systems in terms of their donor masses and radii.

If no spectrum is available, we classify systems as He~CVs if they have $P_{\rm orb}$ of 50--70\,min and undergo dwarf nova outbursts with a recurrence time $\lesssim 3$\,years.
The dwarf nova recurrence times of AM\,CVn-type binaries at such long periods are expected to be decades, even if outbursts occur at all \citep{Levitan2015}.
He~CVs make up the second largest subset of confirmed UCBs in this catalogue (\numhecv\ out of \numconf).

Two systems, CRTS\,J0808+3550 and CRTS\,J1647+4338, have spectra that are very typical of He~CVs but no measured orbital period \citep{Breedt2014}.
We include these as candidate He~CVs.

\textit{sdB donor systems:} This refers to mass-transferring UCBs in which the donor star is spectroscopically confirmed to be a hot subdwarf (sdB-type star).
Two examples are known: ZTF\,J2130+4420 and ZTF\,J2055+4651 \citep{Kupfer2020a,Kupfer2020b}.

\textit{BD donor systems:} This refers to mass-transferring UCBs in which the donor star is a brown dwarf, that is, a dense, hydrogen-rich object with no evidence for helium enhancement relative to typical CVs, but still with $P_{\rm orb} < 70$\,min.
The two known examples are SDSS\,J1507+5230 and WD\,J1540$-$3929.

\textit{Warm donors:} Two systems did not fit neatly into the above categories: ZTF\,J0127+5258 ($P_{\rm orb} = 14$\,min) and \mbox{ATLAS\,J1138$-$5139} ($P_{\rm orb} = 28$\,min).
Both systems have warm donors that are visible in the optical and have optical hydrogen lines, but cannot be spectroscopically classed as hot subdwarfs or brown dwarfs, and both have orbital periods below the range expected for He~CVs.
An evolutionary channel with two phases of common envelope evolution is favoured for both systems.
As these systems appear somewhat similar, we put them into one class. 

\textit{Unclear:} Six confirmed systems (and many candidates) do not have sufficient data available to classify them as one of the above subclasses. 
This was particularly the case where there is no measured \porb\ or observed spectrum; where \porb\ is highly uncertain; or for outbursting systems with  $40 > P_{\rm orb} < 50$\,min, making it difficult to distinguish between classification as an AM\,CVn or a He~CV.
These systems were not classified into any of the above sub-classes in the catalogue, and are marked `Unclear' in Table~\ref{tab:cat3}.

\section{Catalogue Description}

The catalogue in its present form is presented in Tables~\ref{tab:cat1}--\ref{tab:cat4}.
A living version of the catalogue is hosted on the website Zenodo.\footnote{Version 1.0: \url{https://zenodo.org/records/15116793} \newline Latest version: \url{https://doi.org/10.5281/zenodo.8276712}}
Each iteration of the catalogue will be given a unique DOI code and all historic versions will remain accessible on Zenodo.

We now give an overview of the columns contained in the catalogue.
For all relevant columns, the associated uncertainties are listed in a separate column with {\tt \_err} appended to the column name.
References for specific properties are listed in columns with {\tt \_ref} appended, and freeform comments on specific properties are given in columns ending in {\tt \_comment}.

Columns {\tt Name} and {\tt Other\_names} give various names used to refer to each object. 
We have attempted to favour the most commonly-used name in the literature for {\tt Name}.
{\tt RA} and {\tt Dec} give the decimal coordinates, while {\tt Coords\_hex} gives the hexagesimal coordinates.
Coordinates from \textit{Gaia} DR3 were used where available (Section~\ref{sec:gaia}).
{\tt Notes} allows for free-form comments on the target that do not fit into other columns.

We include several boolean flags to help users filter their desired targets from the overall catalogue.
{\tt Confirmed} identifies targets that are confirmed to meet the critera in Section~\ref{sec:criteria}, as opposed to candidates.
{\tt Has\_spectrum} identifies targets for which a spectrum is available, while {\tt Has\_optical\_hydrogen} and {\tt Has\_optical\_donor} identify systems for which the spectrum shows either visible hydrogen lines or spectral lines from the donor star, respectively.
{\tt Has\_eclipses} identifies targets which show partial or complete eclipses of the central white dwarf by the donor star (systems in which only the accretion disc is partially eclipsed are not included).
{\tt Has\_eROSITA} identifies those for which an eROSITA crossmatch has been found. 
We also include boolean columns for several classes of mass-transferring UCB (see Section~\ref{sec:diversity} for further details): {\tt AM\_CVn}, {\tt He\_CV}, {\tt sdB\_donor}, {\tt BD\_donor} (brown dwarf donor), and {\tt Warm\_donor}.

{\tt Discovery\_ref} gives the first reference to classify the target as a UCB.
{\tt Discovery\_year} gives either the year a system was first published, or (for systems with no publication) the year it was first put forward as a candidate.
References to published spectra, where available, are given in {\tt Spec\_ref}.
A cumulative list of references for each target is given in {\tt References}, in addition to references for specific aspects (such as orbital period or masses) that are given in separate columns.
The list of references may not be complete for some well-studied targets, but we have endeavored to include any references which refer to specific breakthroughs for a given target.

The column {\tt Period} gives the orbital period or, where that is not available, the superhump period.
Note that uncertainties on the periods are not tabulated; in most cases these would be \mbox{$< 0.01$\,min}, and the few cases with larger uncertainties are marked in the notes and comments columns.
{\tt Period\_note} gives an addendum to the period: superhump periods are marked \textit{(sh)}, and candidate periods or periods with large uncertainties are marked \textit{?}.
{\tt Period\_method} describes how the period was measured: \textit{Eclipses}, \textit{Spectroscopic}, or \textit{Photometric}.
{\tt Period\_predicted} gives periods predicted based on their outburst properties, following the relationships of \citet{Levitan2015}, only for systems with no measured periods.
{\tt Disk\_state} describes the state of the accretion disc: \textit{High}, \textit{Outburst}, \textit{Low}, or \textit{Direct}. 
The latter value is used in the case of systems with no accretion disc (see the discussion in Section~\ref{sec:accretion}).
{\tt Pdot} gives $dP/dt$ for systems with such measurements, in units of $\times 10^{-12} {\rm s s}^{-1}$. 

For systems with measured mass ratios, these are given in the colum {\tt Q}. 
The column {\tt Q\_method} describes how the mass ratios were derived: \textit{Eclipses}, \textit{Spectroscopy}, \textit{Superhumps}, or \textit{Stage A Superhumps}.
Of these, the label \textit{Superhumps} refers to mass ratios derived using an empirical relation that is observed between the orbital period, superhump period, and mass ratio. 
As discussed in Section~\ref{sec:accretion}, this relation should be used with caution, as it was derived for hydrogen-dominated systems, and is still not well tested for helium accretion discs.
While several forms of the relation have been put forward, we favour the expression of \citet{McAllister2019} due to the inclusion of uncertainties on the relationship, and we recalculate all superhump-derived mass ratios using this expression to ensure homogeneity.
The superhump excess (defined as $\epsilon = [P_{\rm sh} - P_{\rm orb}] / P_{\rm orb}$ for superhump period $P_{\rm sh}$) used is given in column {\tt Superhump\_excess}. 
In principle it is necessary to know the `Stage' (A, B, or C, referring to different parts of an outburst) during which the superhump was measured, but in most sources this is not recorded; if it is not, we assume Stage B, which is the longest-lasting stage of the outburst.
Another method in the {\tt Q\_method} column, \textit{Stage A superhumps}, represents the method proposed by \citet{Kato2013} using superhumps measured during the earliest parts of an outburst; this is a somewhat different method from the \textit{Superhumps} method, and so we do not recalculate these.

Several systems, mostly eclipsing systems, have direct measurements of the component stellar masses $M_1$ (accretor) and $M_2$ (donor). 
These are listed in the columns {\tt M1} and {\tt M2}. 

In several cases where multiple measurements of the same property have been published, we have selected one value as the `canonical' value on a case-by-case basis. 
We favour the purely spectroscopic masses of HM\,Cnc \citep{Roelofs2010} over those of \citet{Munday2023}, which were derived from the $dP/dt$ and are dependent on a greater number of assumptions.
We also favour the purely spectroscopic mass ratio of AM\,CVn itself \citep{Roelofs2006b} over the superhump-derived value, but do include the component masses measured by \citet{Smak2023} even though their derivation relies on the superhump-derived mass ratio.
In each case, we have endeavoured to include all relevant references in the {\tt References} column.

\subsection{Cross-match with \textit{Gaia}}
\label{sec:gaia}

We performed a cross-match of all targets against \textit{Gaia} Data Release 3 \citep[DR3][]{GaiaCollaboration2022}.
For most targets a search radius of 5 arcsec was used. 
For V396\,Hya and GP\,Com, nearby targets with large proper motions, a wider search radius of 20 arcsec was used.
From this crossmatch, we added columns containing the \textit{Gaia} DR3 source IDs ({\tt Gaia\_ID}), magnitudes ({\tt Gaia\_Gmag}), colours ({\tt Gaia\_BPRP}), and parallaxes ({\tt Gaia\_parallax}).
The coordinates in columns {\tt RA} and {\tt Dec} were updated with the \textit{Gaia} coordinates for all targets with a \textit{Gaia} counterpart.
We also took the \textit{Gaia}-based geometric distance estimates of \citet{Bailer-Jones2021}, expressed as a distance median ({\tt Distance}) and the 16th and 84th percentiles ({\tt Distance\_16} and {\tt Distance\_84}).
Of the objects in this catalogue, \numgaia\ out of the \numconf\ confirmed UCBs and \numgaiacand\ out of the \numcand\ candidates have \textit{Gaia} crossmatch results.

\section{The Known Sample}

\begin{figure}
\centering
\includegraphics[width=\hsize]{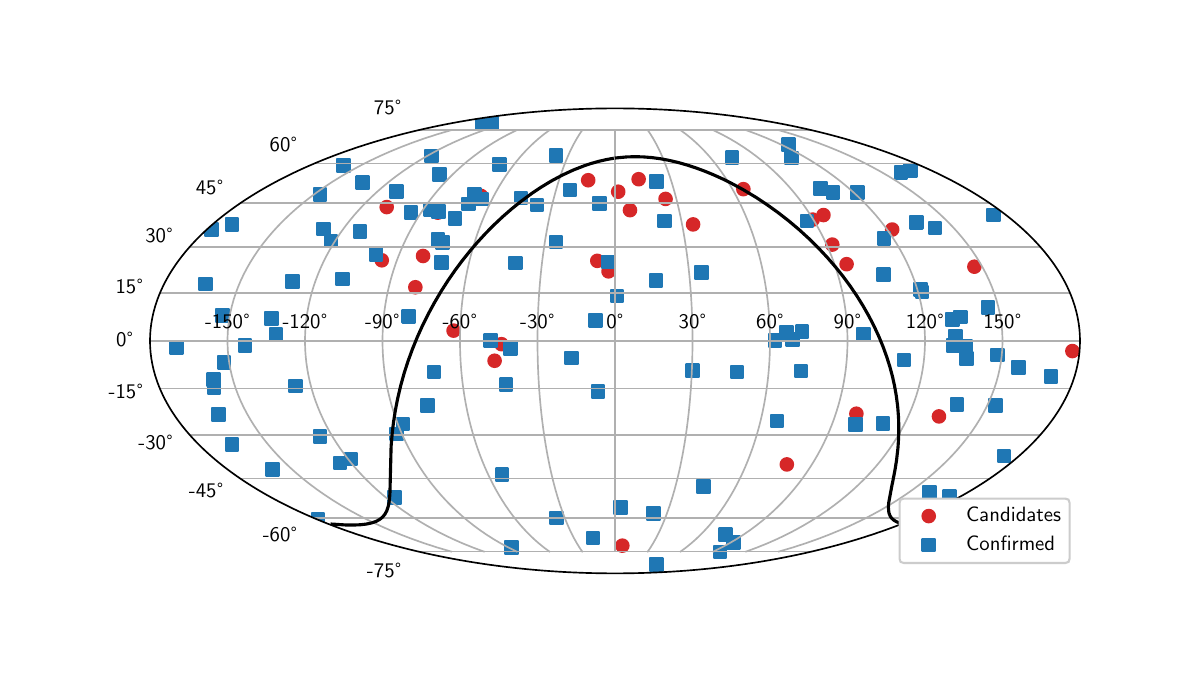}
\caption{The sky location of all known mass-transferring UCBs in equatorial coordinates. 
The solid black line shows the Galactic plane; there is no clear preference for UCBs to be in or out of the plane.
There is a clear bias towards the northern hemisphere, where a greater number of large-sky surveys have been carried out.
}
\label{fig:sky-location}
\end{figure}

Here we provide an overview and summary of the known population of mass-transferring UCBs to date.
Fig.~\ref{fig:sky-location} shows the sky position of targets in the sample.
There is a clear preference for the northern hemisphere, likely because there are more multi-object surveys in the north than in the south.

\subsection{Discovery years and selection effects}
\label{sec:discovery-years}

\begin{figure}
\centering
\includegraphics[width=\hsize]{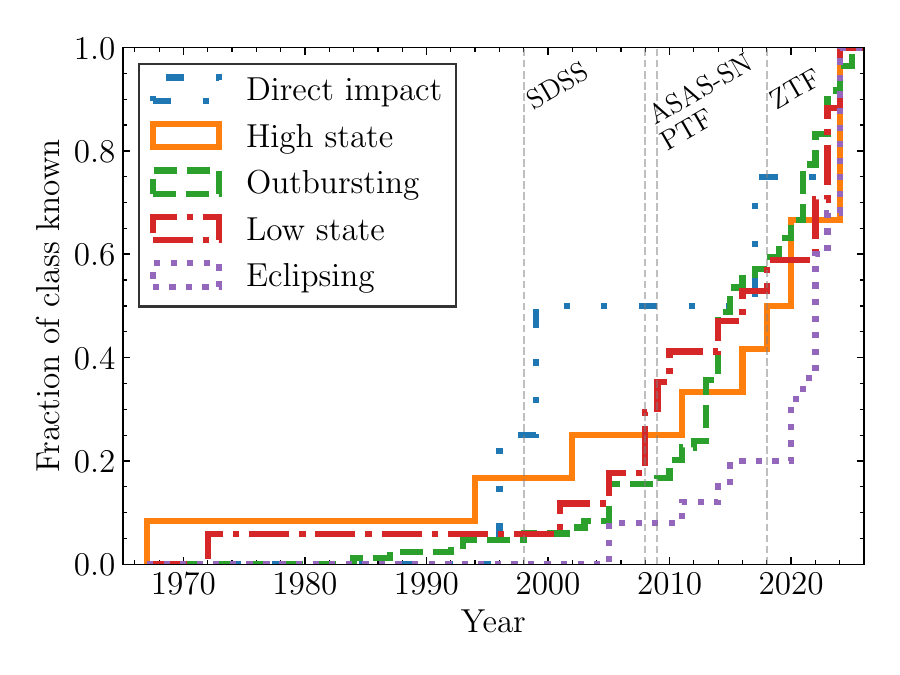}
\caption{Cumulative histograms of the discovery years of several different sub-classes of UCB, highlighting the impact of different surveys on the make-up of the known sample.
SDSS led to an increase in the discoveries of low-state systems (which should dominate in a volume-limited sample), while PTF and ASAS-SN led to rapid discoveries of outbursting systems detected photometrically.
ZTF has led to a sharp rise in the number of eclipsing systems.
}
\label{fig:discovery-year-separate}
\end{figure}

In Fig.~\ref{fig:discovery-year} we have already shown the number of known mass-transferring UCBs by discovery year.
Section~\ref{sec:surveys} discussed the many large-sky surveys that have contributed to the discovery of mass-transferring UCBs.
The impact of various surveys can be seen not only in the number of discoveries, but in the nature of new systems being discovered.
Fig.~\ref{fig:discovery-year-separate} shows cumulative histograms of the known mass-transferring UCBs broken into several sub-classes by accretion disc behaviour and by the presence of eclipses.
The discoveries from new surveys appear $\approx 5-10$\,years after first light.
So there is a rise in the number of low-state systems, which dominate in a magnitude-limited sample, ten years after the beginning of SDSS \citep[e.g.][]{Anderson2005,Anderson2008,Rau2010}, with a smaller bump several years later due to the SDSS-based follow-up survey \citep{Roelofs2009,Carter2013}.
A sharp rise in the number of outbursting systems is visible some years after the beginning of ASAS-SN and PTF \citep[e.g.][]{Levitan2013,Breedt2014}, with a second rise shortly after the beginning of ZTF \citep[e.g.][]{vanRoestel2021}.
It can also be seen that the number of eclipsing systems has increased dramatically in the years since ZTF began operating \citep[e.g.][]{Burdge2020,vanRoestel2022}.
The sharp rise in low-state systems since 2020 is a combination of eclipsing systems from ZTF, X-ray discoveries \citep{Rodriguez2023,Schwope2024a}, and spectroscopic discoveries from SDSS-V \citep{Inight2023a,Inight2023b}.

\subsection{Distances and space density}
\label{sec:distances}

\begin{figure}
\centering
\includegraphics[width=\hsize]{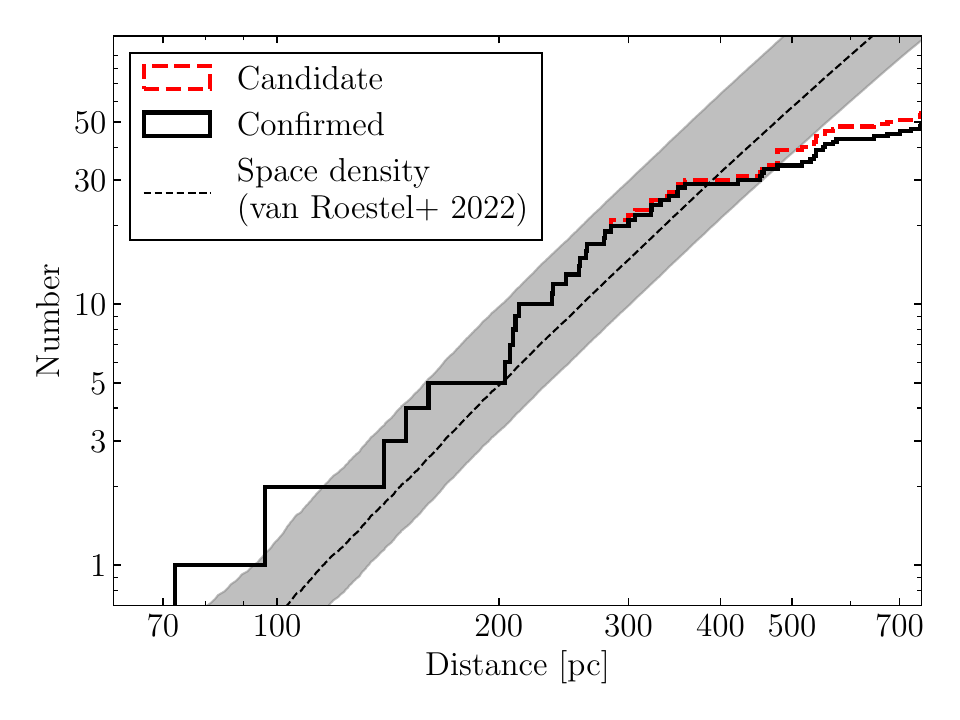}
\caption{
Cumulative histogram of the distances to known mass-transferring UCBs. 
Also plotted is the expected distribution for the local space density measured by \citet{vanRoestel2022}, extrapolated using a Galactic model with a scale height of 300\,pc.
The true systems approximately follow the prediction of \citet{vanRoestel2022} in the range 150--300\,pc, after which incompleteness becomes much more severe.
There appears to be an excess at short distances when compared to the prediction (two systems within 100\,pc where only one would be predicted), but given the small numbers this is not statistically significant.
}
\label{fig:distances}
\end{figure}

To date, the only population-level statistic that has been measured for the mass-transferring UCB population is the space density.
Several groups have estimated this for the hydrogen-depleted AM\,CVn-type systems and found consistent results, though always with small sample sizes.
\citet{Carter2013} used a magnitude-limited sample of four AM\,CVn binary systems in the SDSS-III footprint to estimate the space density of AM\,CVn-type binaries at $\rho = (5 \pm 3) \times 10^{-7}$\,pc$^{-3}$.
\citet{Ramsay2018} used the \textit{Gaia} distances of known AM\,CVn binary systems to estimate a lower limit, $\rho > 7 \times 10^{-8}$\,pc$^{-3}$.
Based on the seven eclipsing AM\,CVn-type systems recovered from ZTF, \citet{vanRoestel2022} found $\rho = 6^{+6}_{-2} \times 10^{-7}$\,pc$^{-3}$.
\citet{Rodriguez2024} used the sample of three AM\,CVn-type systems recovered from the western eROSITA dataset within 150\,pc to determine a space density of $\rho = (5.5 \pm 3.7) \times 10^{-7}$\,pc$^{-3}$.

In Fig.~\ref{fig:distances} we plot geometric distance estimates \citep{Bailer-Jones2021} for nearby mass-transferring UCBs.
We also plot, for comparison, a predicted distribution based on the space density estimate of \citet{vanRoestel2022}.
The distribution was calculated using the method of \citet{Pretorius2007}, assuming a Galactic disc with a scale height 300\,pc, which is consistent with the best fit for CVs \citep{Pala2020}.

It appears that our distance distribution is consistent with the \citet{vanRoestel2022} density estimate and follows the expected Galactic distribution out to $\approx 300$\,pc.
This suggests that our completeness is relatively high within 300\,pc and falls significantly beyond that limit.
Note, however, that systems within 300\,pc have continued to be discovered recently \citep{vanRoestel2022,Inight2023b,Rodriguez2023}.
When we consider also that the southern hemisphere remains under-explored relative to the northern hemisphere (Fig.~\ref{fig:sky-location}), it seems likely that more systems are waiting to be discovered within 300\,pc.

\subsection{Orbital periods}
\label{sec:periods}

Orbital period is the most widely measured physical property of the UCBs in this catalogue: \numporb\ out of the \numconf\ confirmed UCBs have a measured orbital or superhump period.
The most common method by which orbital periods are determined is photometric measurement (42 systems), followed by spectroscopic measurement (33 systems) and timing of eclipses (25 systems; this includes systems with partial disc eclipses, even when the central white dwarf is not eclipsed).

In Fig.~\ref{fig:periods} we showed the period distribution of the known systems.
When considering all known systems, the period distribution is strongly affected by the selection biases discussed in Section~\ref{sec:discovery-years}; most notably, the prominent peak of outbursting systems, which are preferentially found through photometric variability surveys.
In principle it should be possible to study the underlying period distribution of the population while modelling the selection effects of defined sub-sets of the sample, such as volume-limited or eclipsing sub-samples.

\subsection{Magnitudes and colours}

\begin{figure}
\centering
\includegraphics[width=\hsize]{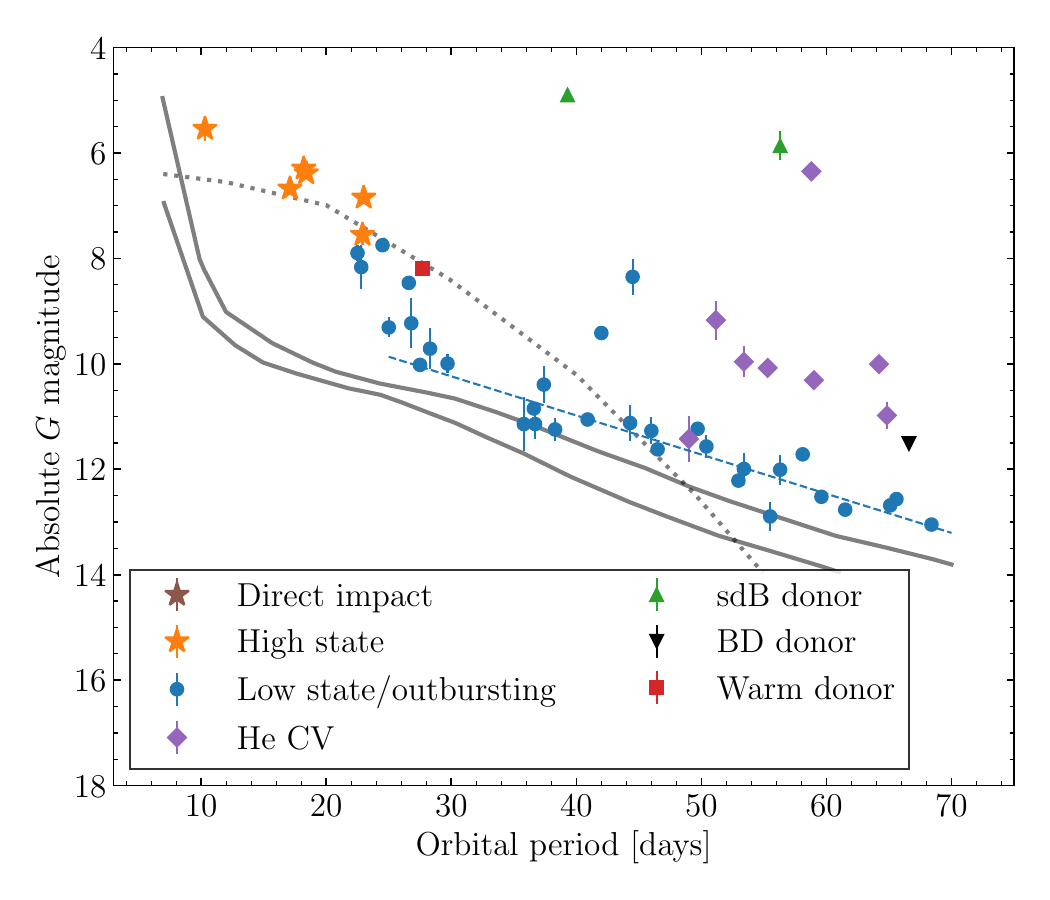}
\caption{Absolute magnitudes of targets in this catalogue as a function of orbital period. We have only included targets for which the \textit{Gaia} parallax is detected with $5 \sigma$ confidence.
The grey solid lines show the white dwarf magnitudes predicted by \citet[][]{Bildsten2006} for white dwarf donors of two different levels of initial entropy, and the grey dotted line shows the accretion disc magnitude predicted by \citet{Nelemans2004}.
AM\,CVn-type systems in the low state follow an approximately linear relationship with orbital period (fitted by the blue dashed line).
Other types of systems are generally brighter, due to extra light from one or more of the donor star or the accretion disc.
As was previously discussed by \citet{Ramsay2018}, even low-state systems at long orbital periods are systematically brighter than the model predictions.
}
\label{fig:magnitudes}
\end{figure}

\citet{Ramsay2018} noted that the absolute magnitude of AM\,CVn-type systems depends strongly on their orbital period, reflecting the correlation between mass transfer rate and orbital period and the resulting changes in accretion behaviour described in Section~\ref{sec:accretion}.
In Fig.~\ref{fig:magnitudes} we show the absolute $G$-band magnitudes of our sample as a function of orbital period, using distances from \citet{Bailer-Jones2021}.
We have removed any targets for which the \textit{Gaia} parallax is detected with less than $5 \sigma$ significance.
We also plot the expected magnitude of the accretion disc \citep{Nelemans2004} and the accreting white dwarf with an assumed white dwarf donor and two possible mass transfer rates \citep{Bildsten2006}. 
Both of these theoretical predictions have been converted from the $V$ to $G$ pass-bands\footnote{\url{https://gea.esac.esa.int/archive/documentation/GDR2/Data\_processing/chap\_cu5pho/sec\_cu5pho\_calibr/ssec\_cu5pho\_PhotTransf.html}} using the median $BP-RP$ colour of AM\,CVn-type binaries, but the correction was only 0.02 magnitudes.

Several patterns can be observed among the classical AM\,CVn-type systems.
Firstly, the brightest systems are high-state and direct-impact systems (yellow and brown stars) with $P_\mathrm{orb} \lesssim 20$\,min, where the emission is dominated by accretion luminosity.
At $P_{\rm orb} \gtrsim 25$\,min the accretor-dominated AM\,CVn-type systems (blue circles) generally follow the \citet{Bildsten2006} prediction for the accretor luminosity, but are slightly elevated.
This is consistent with the finding that estimates of accretor temperature are elevated above model-derived predictions \citep{Macrie2024}.
This elevation may be the result of their non-DB atmospheres or of the low donor masses, and hence low donor entropy and low mass transfer rates, assumed by \citet{Bildsten2006}.
Recently it has also been suggested that additional orbital angular momentum loss, such as magnetic braking, may speed up the period evolution of AM\,CVn-type binaries \citep{Belloni2024,Maccarone2024}, which may then affect the relation between orbital period and accretor luminosity. 
In the 20--25\,min period range, frequent outbursts may bias the measured magnitudes upwards, resulting in the apparent `transition' region between the low state and high state systems.
\citet{Ramsay2018} gave a detailed discussion of individual AM\,CVn-type systems on this plot, including the notable outliers ASASSN-14ei ($P_\mathrm{orb} = 42$\,min) and SDSS\,J0804+1616 (44.5\,min).

Other types of mass-transferring UCBs do not follow these same patterns.
The two sdB donor systems \citep{Kupfer2020a,Kupfer2020b} are dominated by their donor luminosity and so are substantially brighter.
The He~CVs (purple diamonds) are generally brighter than classical AM\,CVns at the same orbital period, due to their higher mass transfer rates and (in some cases) the contribution of light from their donor stars.
Donor contribution is also significant in the magnitude of ATLAS\,J1138-5139 8 (a warm-donor binary with $P_\mathrm{orb} = 28$\,min).
SDSS\,J1507+5230 (a brown dwarf donor, $P_\mathrm{orb} = 67$\,min) is also brighter than an AM\,CVn-type binary at that orbital period, but has no measurable contribution in the optical from its donor star \citep{Littlefair2007}, and so its brightness is likely driven by a higher mass transfer rate.

For future analyses, it may be useful to parametrise the dependence of absolute magnitude on orbital period. 
For low-state or outbursting classical AM\,CVn-type systems (blue circles), the dependence is well described (for orbital periods of 25--70\,min) by 
\begin{equation}
M_G \left( P_{\rm orb} \right) = 7.9(2) + 0.076(4) \frac{ P_{\rm orb}}{\rm min} ~ {\rm mag}.
\label{eq:mag-periods}
\end{equation}
Numbers in parentheses are uncertainties on the last digit of the corresponding value.
To arrive at this fit we performed a sigma-clipping routine on the low state AM\,CVn-type systems, iteratively removing the most significant outlier and repeating the fit until all remaining points were within $3 \sigma$ of the best-fit line.
This best-fit line is plotted as a blue dashed line in Fig.~\ref{fig:magnitudes}.

\begin{figure}
\centering
\includegraphics[width=\hsize]{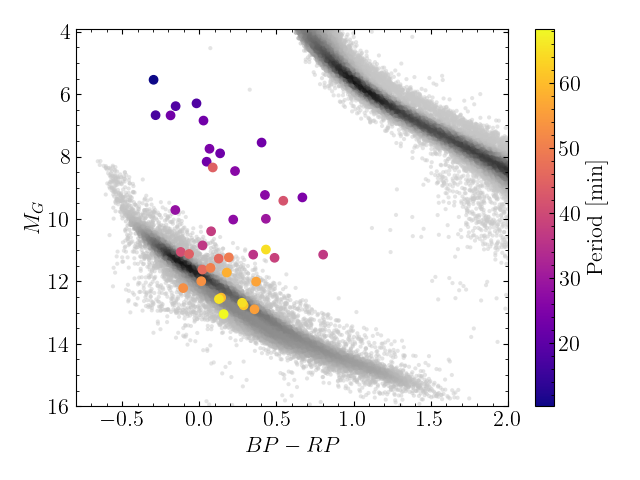}
\caption{Colour-magnitude diagram of confirmed AM\,CVn-type binaries in this catalogue.
Points are coloured according to their orbital periods.
Also plotted are volume-limited samples of main sequence stars and white dwarfs, drawn randomly from \textit{Gaia} DR3.
The AM\,CVn-type systems follow a `Z' pattern pointed out by \citet{vanRoestel2022}: short period systems are elevated above the white dwarf sequence because they are dominated by the accretion disc, while longer-period systems follow a white dwarf cooling sequence.
}
\label{fig:cmd}
\end{figure}

In Fig.~\ref{fig:cmd}, we plot the \textit{Gaia} colour-magnitude diagram of the confirmed AM\,CVn-type (completely hydrogen depleted) binary systems.
A `Z'-shaped dependence of colour and magnitude on orbital period was noted by \citet{vanRoestel2022}.
Disc-dominated systems ($P_\mathrm{orb} \lesssim 20$\,min) are generally bluest at the shortest periods, and become redder towards $P_\mathrm{orb} \approx 20$\,min.
Then, as they transition to become accretor-dominated, they become bluer again, until they merge with the white dwarf sequence at $P_\mathrm{orb} \approx 30$--40\,min.
Towards $P_\mathrm{orb} \gtrsim 40$\,min the systems cool and redden in parallel with white dwarf cooling tracks.

\begin{figure}
\centering
\includegraphics[width=\hsize]{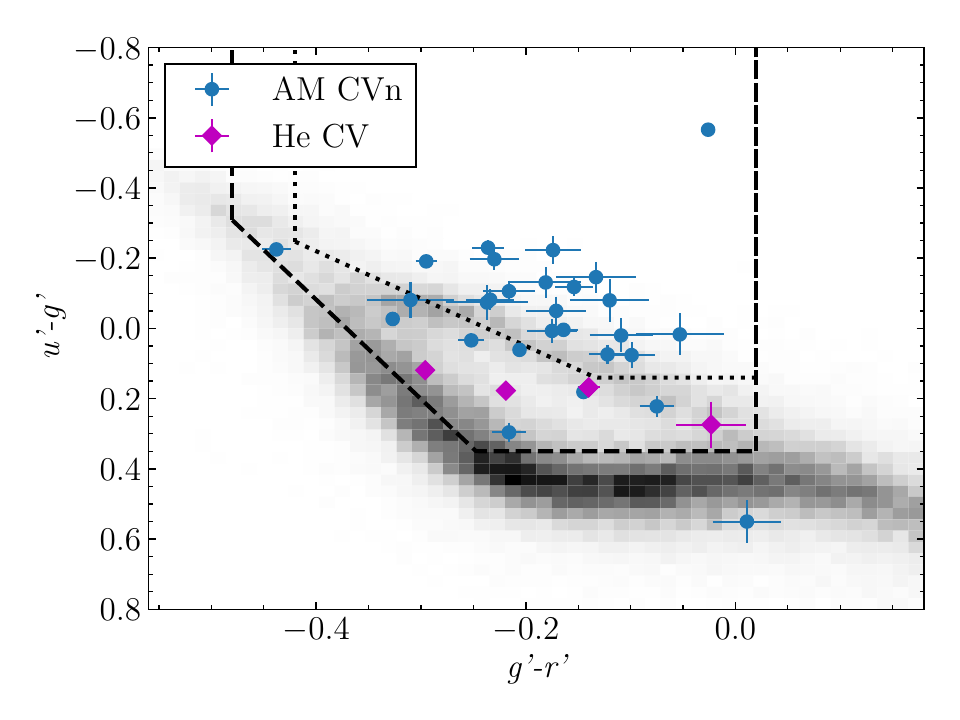}
\caption{Colour-colour plot of targets in this sample with SDSS $u'g'r'$ colours and $g' < 20.5$. A further four points (two AM\,CVn-type and two He~CVs) are not shown because they would be outside the panel. The grey two-dimensional histogram shows the distribution of a sample of white dwarfs \citep{GentileFusillo2021a}. The black dotted line shows the original colour-colour selection box of \citet{Roelofs2009}, and the black dashed line shows the updated selection criteria of \citet{Carter2013}. A number of mass-transferring UCBs are missed by the older selection criteria but selected by the newer criteria.
}
\label{fig:colours}
\end{figure}

\citet{Roelofs2009} and \citet{Carter2013} used colour selection to identify candidate UCBs for spectroscopic follow-up.
Ten of the targets in this catalogue (nine AM\,CVn-type systems and one He~CV) were found by that survey, making it a significant contributer to the known sample.
The colour cuts were defined by \citet{Roelofs2009} to include all AM\,CVn-type binaries known at that time, and the selection was assumed to be complete in terms of colour by \citet{Carter2013} for the space density analysis discussed in \ref{sec:distances}.
The discovery of several AM\,CVn-type systems outside the original selection box led \citet{Carter2014b} to propose an expanded set of selection criteria, but follow-up spectroscopy was not obtained using the new criteria.

In Fig.~\ref{fig:colours}, we plot the SDSS $u'g'r'i'$ colours for all targets in this catalogue within a limiting magnitude of $g' < 20.5$, using the cross-match performed by \citet{GentileFusillo2021a}.
We also plot the selection boxes of \citet{Roelofs2009} and \citet{Carter2014b}.
Of the known AM\,CVn-type systems, 29 have SDSS $u'g'r'i'$ (not including a further two which are outside the limits of this plot, presumably due to unreliable colour measurements or contamination from nearby stars).
25 of those are selected by the updated criteria of \citet{Carter2014b}, while only 17 are selected by the original cuts of \citet{Roelofs2009} that were implemented in the follow-up survey.
While not the intended target of that survey, several He~CVs are selected by the updated cuts while none are selected by the original cuts.
It should therefore be emphasized that future colour-based searches for mass-transferring UCBs should use the \citet{Carter2014b} criteria or similar, rather than the more conservative \citet{Roelofs2009} criteria.

\subsection{Masses and radii}
\label{sec:masses}

\begin{figure}
\centering
\includegraphics[width=\hsize]{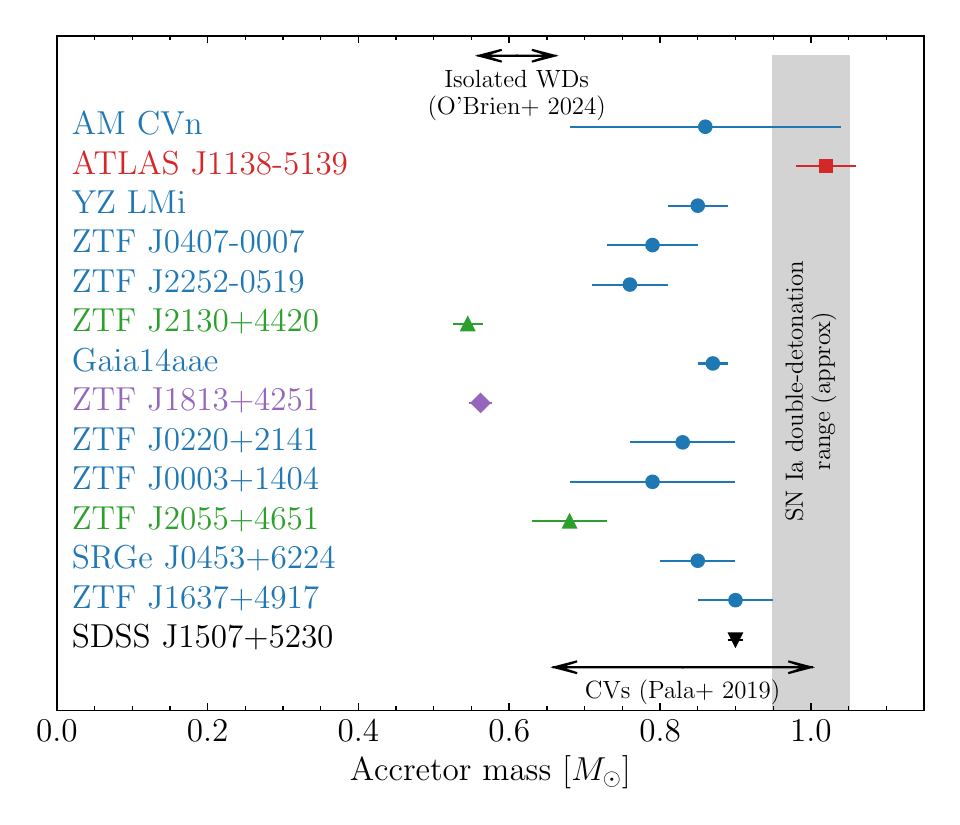}
\caption{The measured masses of the accreting components of the binaries in this catalogue.
The marker shapes and colours match those in Fig.~\ref{fig:magnitudes}.
All masses plotted here were measured by eclipses, except AM\,CVn itself which is measured spectroscopically \citep{Smak2023}.
Also labelled are the typical masses for white dwarfs in isolation and in mass-transferring binaries, and the approximate mass range for which typical Type~Ia supernovae are expected.
We note that ATLAS\,J1138$-$5139 is the only binary with sufficient mass for a typical Type~Ia supernova.
The two accretors with sdB-type donors (green carets) and the pre-bounce He~CV (magenta diamond) are lower in mass than the rest of the distribution, presumably because their accumulated mass from accretion is still low; their masses are more typical of isolated white dwarfs, while the majority of systems have masses more typical of accreting white dwarfs.
}
\label{fig:m1s}
\end{figure}

\begin{figure*}
\centering
\includegraphics[width=0.49\hsize]{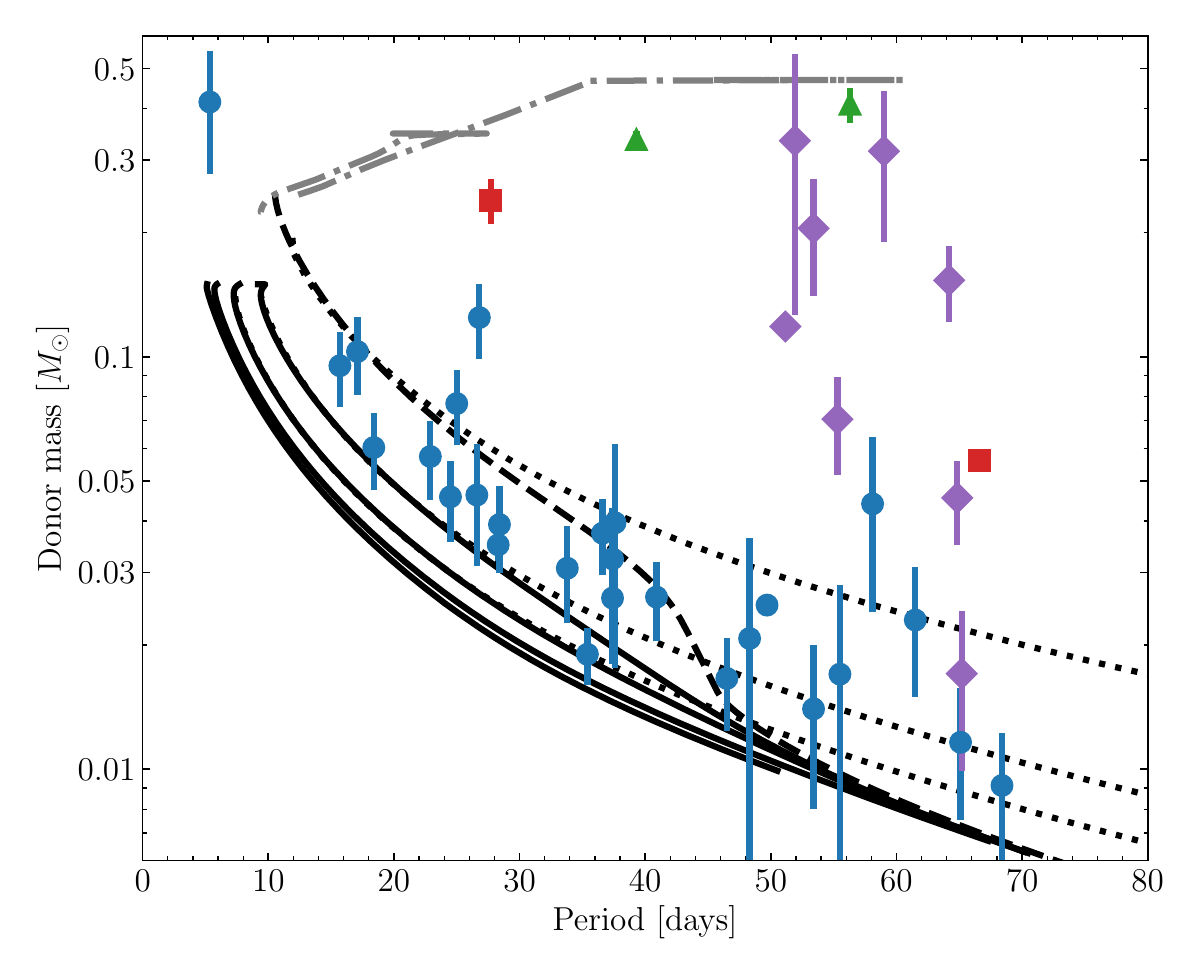}
\includegraphics[width=0.49\hsize]{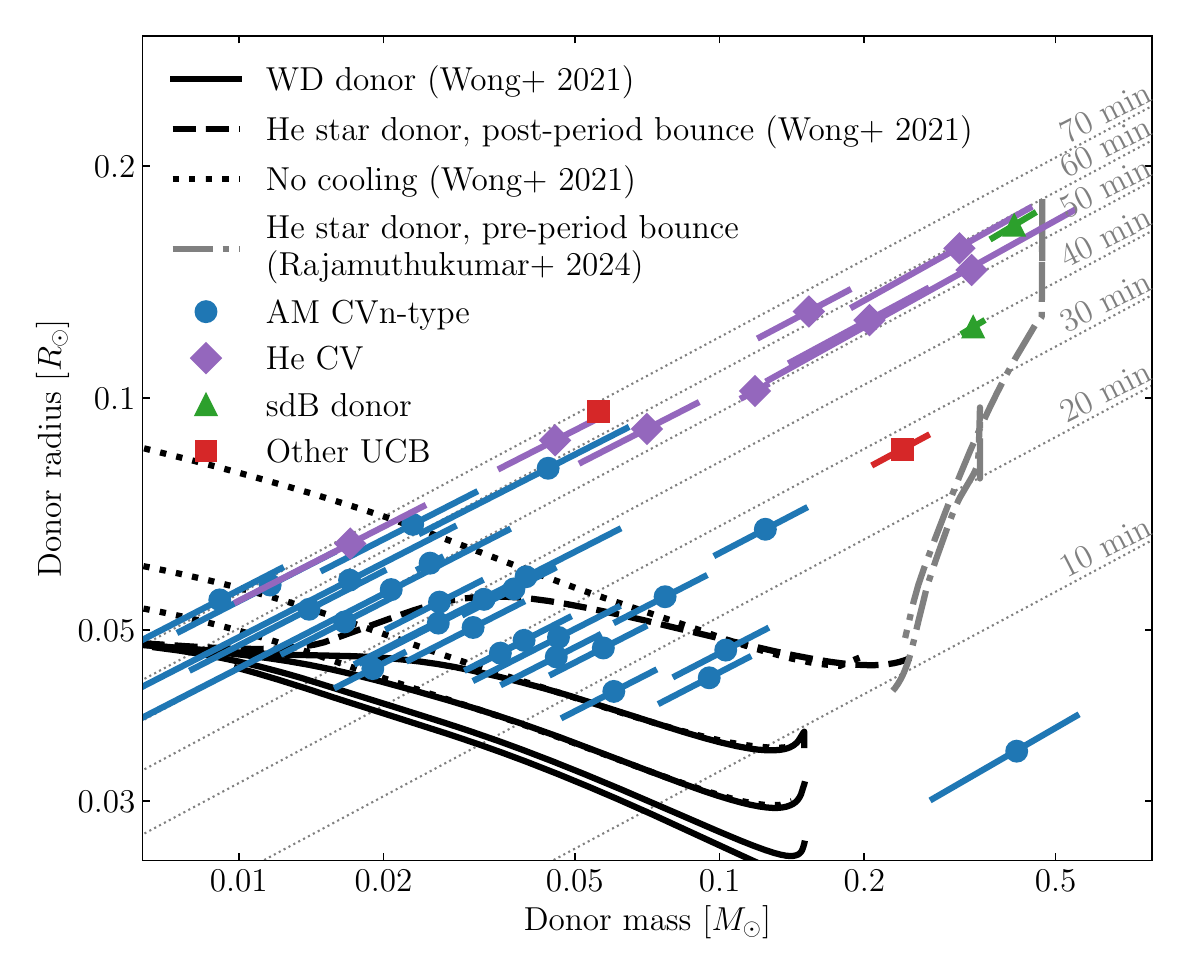}
\includegraphics[width=0.49\hsize]{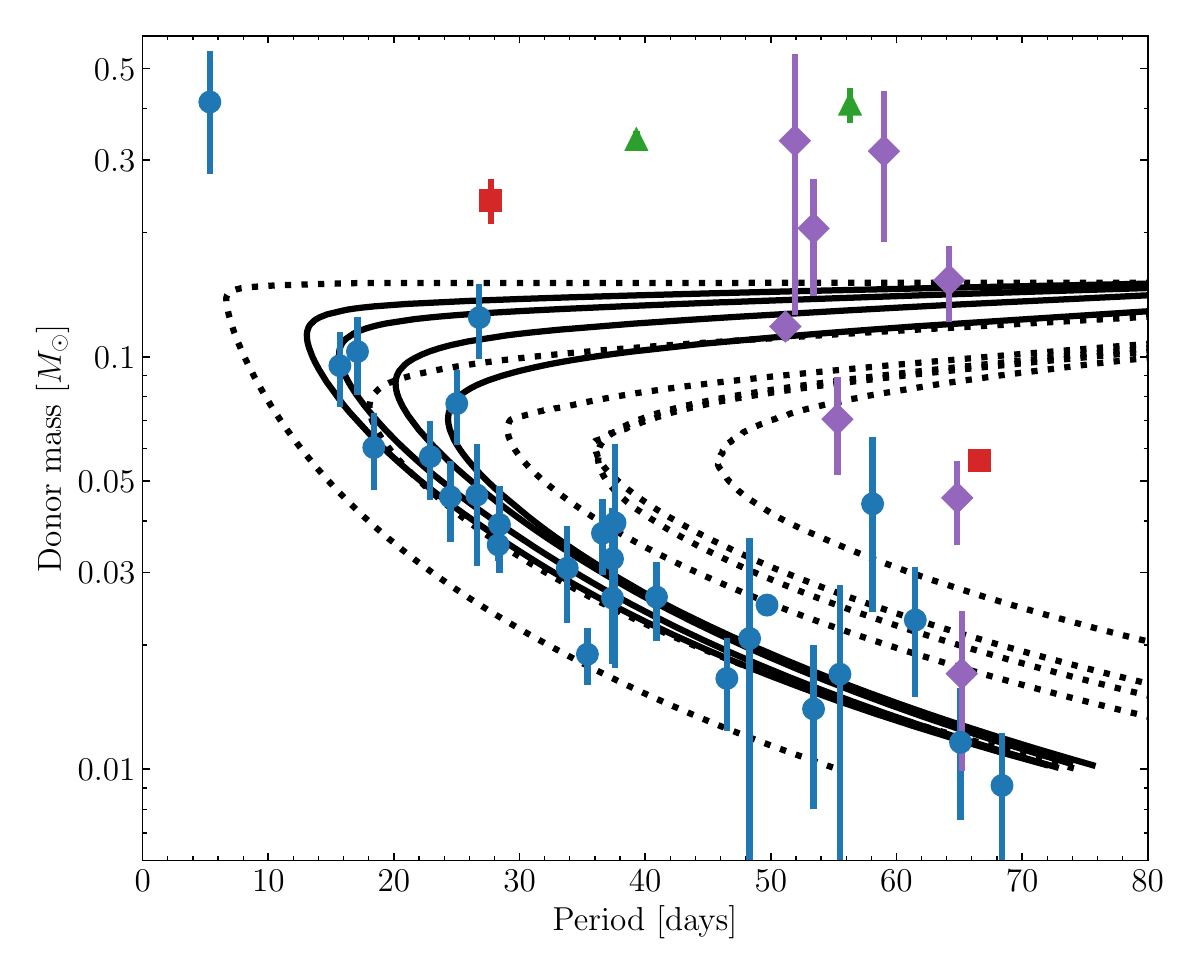}
\includegraphics[width=0.49\hsize]{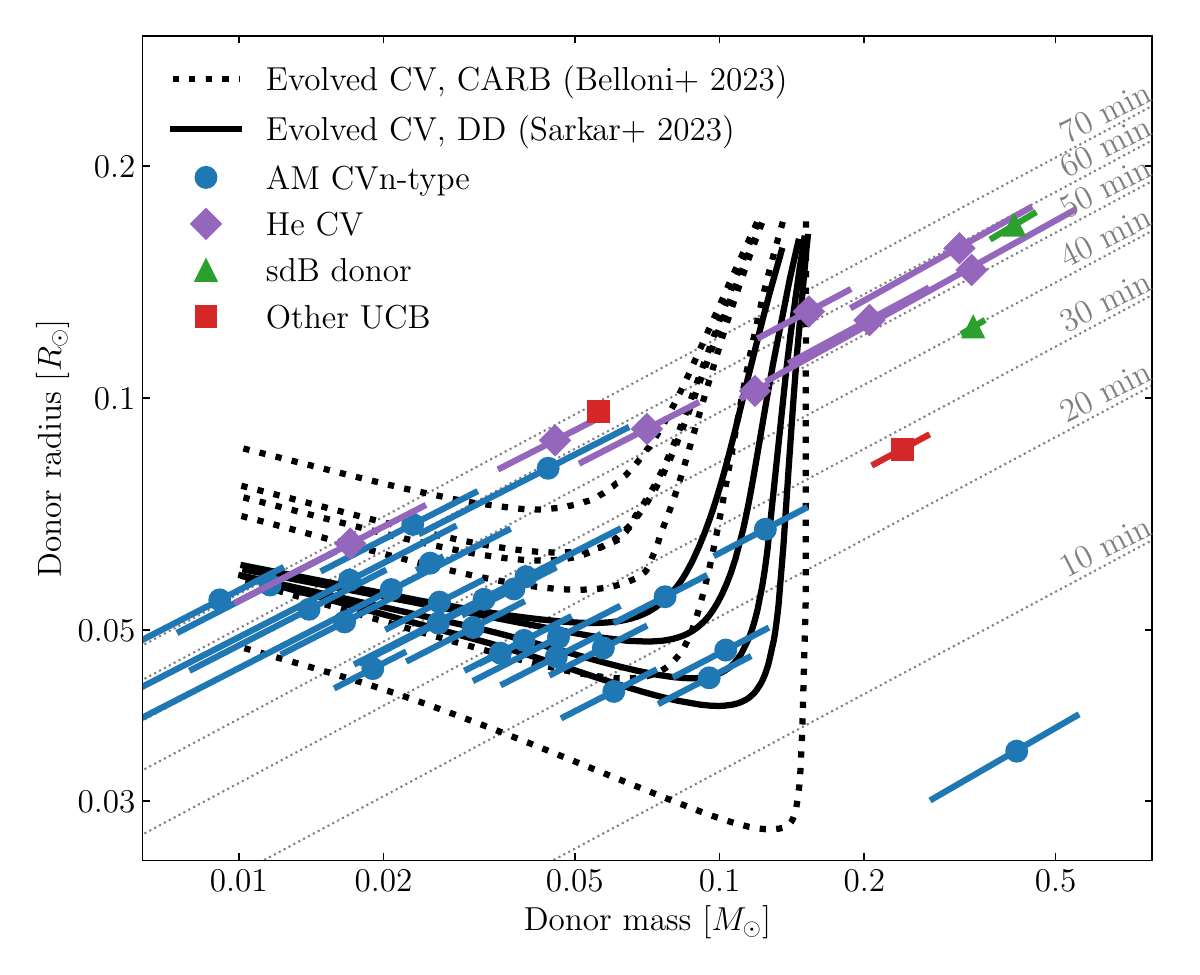}
\caption{The measured orbital periods and donor star masses and radii for mass-transferring UCBs in the catalogue. 
In the left panels, we plot the observed properties: orbital period and donor mass.
In the right panels, we plot donor mass and radius.
Uncertainties in the right panels are diagonal because of the strong constraint on donor density that arises from the orbital period.
The two ``other'' systems plotted are the brown dwarf donor SDSS\,J1507+5230 and the 30\,min supernova progenitor candidate ATLAS\,J1138-5139.
We also plot several evolutionary tracks from different scenarios, as discussed in the text: the upper panels show evolutionary models of the white dwarf and helium star donor tracks, while the lower panels show evolutionary models from the evolved CV scenario.
The systems without hydrogen (AM\,CVn-type binaries) follow a sequence from short to long orbital periods, while He~CVs with hydrogen follow a different sequence from low to high masses.
In general, any formation scenario can explain at least some fraction of the known systems, and no single formation scenario can explain every system.
}
\label{fig:donor-mr}
\end{figure*}

A minority of mass-transferring UCBs have measured stellar masses and radii, primarily measured by eclipse light curve fitting.
In Fig.~\ref{fig:m1s}, we plot the measured masses of accretors across all systems in this catalogue.
All of these accretor masses are measured from eclipses, except for the spectroscopically-derived mass of AM\,CVn itself \citep{Smak2023}.

The majority of accretor masses are consistent with the typical $0.83 \pm 0.17 M_\odot$ mass range observed across accreting white dwarfs in CVs \citep{Pala2020}.
As discussed in Section~\ref{sec:gwsn}, ATLAS\,J1138$-$5139 has an unusually high-mass accretor, and is a promising candidate for a Type~Ia supernova progenitor.
The two systems with sdB donor stars, ZTF\,J2130+4420 and ZTF\,J2055+4651, are more consistent with the typical $\approx 0.6 M_\odot$ masses seen in non-accreting white dwarfs \citep[e.g.][]{OBrien2024}.
These systems are thought to be in an early stage of mass transfer \citep{Bauer2021} and have likely not yet accreted a significant helium envelope.
ZTF\,J1813+4251, a He~CV which is thought to be passing through a transitional stage between the longer-period CVs and the AM\,CVn-type binaries \citep{Burdge2022}, also has a low-mass accretor more typical of a non-accreting white dwarf. 

In Fig.~\ref{fig:donor-mr}, we plot all of the UCBs that have measured donor masses.
In the left column we plot measured units ($P_{\rm orb}$ and $M_2$), while in the right column we plot $M_2$ and $R_2$, following the plots of \citet{SolheimAMCVn}, \citet{Green2018b,Green2018a,Green2020}, and \citet{vanRoestel2022}.
There is a well-established relationship between the orbital period and mean density of a Roche-lobe filling donor star \citep{Faulkner1972a}.
As the period is tightly constrained for all systems on this plot, the uncertainties on $M_2$ and $R_2$ are strongly correlated, and we plot the error bars diagonally along lines of constant period to show this.

There are many more systems with measurements of mass ratio ($q = M_2 / M_1$) than with direct measurements of $M_2$.
Where only $q$ is measured, we adopt $M_1 = 0.83 \pm 0.17$ \citep{Pala2020}, and estimate $M_2$ in this way.
As discussed in previous sections, in the case of systems for which $q$ was derived from Stage B or C superhumps, we have re-derived $q$ using the relations of \citet{McAllister2019} to ensure consistency.

We also plot evolutionary tracks from several different sources.
In the upper panels, we plot white dwarf and helium star donor models.
For the white dwarf donor channel, we use the models of \citet{Wong2021}.
The vertical distribution of tracks corresponds to white dwarf donors with varying levels of thermal inflation of the donor at the onset of mass transfer.
Binaries with wider separations at the end of the common envelope phase have longer inspiral times preceding the onset of mass transfer and hence more time for the donor to cool, resulting in more compact donor stars.
Because of the uncertainties outlined in that paper about the subsequent cooling rate of the donor star during mass transfer, we plot two sets of tracks: one set with nominal donor cooling rates, and a further set of tracks with equivalent initial conditions but for which donor cooling was not allowed.
The latter case causes donors to remain at higher levels of inflation as they evolve towards longer orbital periods, which helps to explain the high-mass and large-radius donors observed at longer periods.

We also plot the helium star donor track modelled by \citet{Wong2021}.
Post-bounce, helium star donors move approximately parallel to white dwarf donors in mass-radius space, with larger radii for a given mass.
To demonstrate the pre-bounce evolution of the helium star channel and the connection to the two sdB-donor systems, we plot two evolutionary tracks for sdB-donor binary systems from \citet{Rajamuthukumar2024}.
The models of \citet{Rajamuthukumar2024} were terminated at the first helium nova eruption, due to uncertainties around the modeling of nova.

In the lower panels, we plot models of the evolved CV formation scenario, using models with enhanced magnetic braking by \citet{Belloni2023} and \citet{Sarkar2023}.
Although the adopted prescriptions for magnetic braking are different between the two works , the distributions of the tracks are qualitatively similar.
Earlier evolved CV models without enhanced magnetic braking \citep[e.g.][]{Goliasch2015} were generally limited to $P_{\rm orb} \gtrsim 40$\,min and more massive donors.

The AM\,CVn-type systems (systems with no detected hydrogen; the blue data points in Fig.~\ref{fig:donor-mr}) form a sequence in donor mass-period or mass-radius space.
Under the white dwarf or helium star donor channels, this is an evolutionary sequence, running parallel with the evolutionary tracks of \citet{Wong2021} from short to long orbital periods.
In evolved CV models with enhanced magnetic braking \citep{Belloni2024,Sarkar2023}, the AM\,CVn sequence is again an evolutionary sequence which systems follow after going through a period bounce, with more inflated systems generally having undergone the period bounce at longer orbital periods.

He~CVs (the magenta points in Fig.~\ref{fig:donor-mr}) do not follow the same sequence as classical AM\,CVn binaries, but span a range of donor masses with periods always in the range of 50--70\,min.
At periods of 50--70 minutes and $M_2 \lesssim 0.05 M_\odot$, the AM\,CVn and He~CV regions of parameter space overlap.
Under the evolved CV scenario, some or all of these He~CVs can be understood as progenitors of the hydrogen-depleted AM\,CVn-type systems.

Each set of model tracks is consistent with at least some observed systems, and most observed systems can be explained by multiple evolutionary scenarios.
Tracks from the white dwarf or helium star donor scenarios can generally explain most of the classic AM\,CVn-type systems as well as the sdB-donor binaries and other binaries with warm donors at short orbital periods (e.g. ZTF J0127+5258 and ATLAS J1138$-$5139), but still struggle to explain some of the more inflated donors, especially at longer periods, and cannot explain most He~CVs.
Evolutionary tracks from the evolved CV scenario can again explain most of the AM\,CVn-type systems, as well as most He~CVs, but struggle to explain some high-mass donors at short periods.
It seems likely that there are two or three evolutionary scenarios at play in the formation of the AM\,CVn-type binaries.

In most cases it is not possible to determine for any given system which formation channel it must have descended from, as the measurable properties that can arise from different channels have a significant degree of overlap.
Instead, the most promising avenue to test the proposed formation channels is on a population level, comparing observed population statistics such as space density, or the distribution of orbital periods and masses, against those predicted by different channels.

\section{Summary}

We have presented a catalogue of mass-transferring UCBs, including all known binary systems with orbital periods $< 70$\,min containing accreting white dwarfs.
The catalogue is presented in the tables in Appendix~\ref{sec:appendix-catalogue}.
A living version of the catalogue will be hosted on the Zenodo platform and updated regularly.

Given their compact orbits, these binaries necessarily consist of two evolved and dense objects.
In almost all cases the accreted material is partially or completely hydrogen-depleted.
The majority of systems in the catalogue belong to the class of systems called AM\,CVn-type binaries, which we define as having no detectable hydrogen, while a minority belong to the partially hydrogen-depleted class called He~CVs.
The two classes are thought to be evolutionarily related under some formation scenarios, and occupy similar but distinct regions of parameter space in terms of orbital periods, outburst behaviours, masses and radii.
Several other unusual binary systems also meet the criteria for inclusion in the catalogue.
The literature overview in Section~\ref{sec:overview} may be a useful resource for those who are new to the field.

We have given an overview of a number of observable properties of the known sample of mass-transferring UCBs.
The sample shows severe incompleteness effects at distances $\gtrsim 300$\,pc, but within 300\,pc agrees well with previous observational space density estimates.
The distribution of orbital periods shows strong biases, particularly towards period ranges that exhibit dwarf nova outbursts.

We have presented an overview of the stellar masses and radii measured to date.
In particular, the donor masses and radii can give insight into the formation history of a binary system.
Of the three proposed formation channels, evolutionary tracks for all three could explain most of the population, and no channel can explain every mass-transferring UCB.
In order to better understand the formation history of these systems, and the relative importance of the proposed formation channels, a population-level comparison between models and observed systems may be the next step for the field.
This will require effort on both the observational front (continuing to characterise the growing sample of known systems) and the theoretical (predicting population-level distributions of parameters that can be compared to the systems in this catalogue).

\begin{acknowledgements}

This paper is dedicated to the memory of Tom Marsh, a mentor, colleague and friend.
The authors are grateful to the organisers of the 2022 Kavli Institute of Theoretical Physics workshop on white dwarfs, where this paper was first planned, and to the organisers of the 2023 AM\,CVn 5 workshop at Armagh Observatory, where many useful discussions took place.
The authors would like to thank the anonymous referee for their helpful comments and feedback.
We are grateful to Evan Bauer, Diogo Belloni, Anton Biryukov, Boris G\"{a}nsicke, Stuart Littlefair, Gavin Ramsay, Liliana Rivera-Sandoval, Matthias Schreiber, Tony Rodriguez, and Arnab Sarkar for helpful comments and discussions.
We are also grateful to James Munday and Anwesha Sahu for providing feedback on early versions of the catalogue.
MJG thanks Mitch Begelman and the JILA department at the University of Colorado, Boulder, for generously providing office space at which much of this paper was written.
MJG acknowledges support from the European Research Council through ERC Advanced Grant No. 101054731.

This work made use of the software packages {\sc python, numpy, matplotlib, scipy, astropy, emcee,} and {\sc trm}.

\end{acknowledgements}

%
%


  \bibliographystyle{aa} 
  \bibliography{refs} 





   
  



\begin{appendix}

\section{Catalogue}
\label{sec:appendix-catalogue}

Tables~\ref{tab:cat1}--\ref{tab:cat4} contain a simplified overview of the catalogue of all known mass-transferring UCBs.
The full catalogue, including many columns not shown here, can be found in the online material for this paper, and on Zenodo at \url{https://doi.org/10.5281/zenodo.8276712}.

\begin{table*}
\caption{List of confirmed AM\,CVn-type (without detected hydrogen) systems in the catalogue.
\label{tab:cat1}
}
\begin{tabular}{lcccccc}
\hline
Name & Coordinates & $P_{\rm orb}$ & Disc state & $G$ mag. & $q$ & Has \\
& & [min] &&& & spectrum \\
\hline
HM Cnc & 08:06:22.84 +15:27:31.5 & 5.35 & Direct & $20.92 \pm 0.03$ & $0.50 \pm 0.13$ & \checkmark \\
eRASSU J0608-7040 & 06:08:39.5 -70:40:14 & 6.20 & Direct & -- & -- & $\times$ \\
ZTF J0546+3843 & 05:46:27.41 +38:43:13.4 & 7.95 & High & $19.31 \pm 0.01$ & -- & \checkmark \\
ZTF J1858-2024 & 18:58:05.95 -20:24:48.6 & 8.68 & High & $19.37 \pm 0.01$ & -- & \checkmark \\
V407 Vul & 19:14:26.09 +24:56:44.6 & 9.48 & Direct & $19.36 \pm 0.00$ & -- & \checkmark \\
ES Cet & 02:00:52.17 -09:24:31.7 & 10.3 & High & $16.80 \pm 0.00$ & -- & \checkmark \\
ZTF J0425+3858 & 04:25:50.21 +02:35:55.8 & 13.2 & High & -- & -- & \checkmark \\
SDSS J1351-0643 & 13:51:54.46 -06:43:09.0 & 15.7 & High & $18.72 \pm 0.00$ & $0.11 \pm 0.00$ & \checkmark \\
AM CVn & 12:34:54.60 +37:37:44.1 & 17.1 & High & $14.06 \pm 0.00$ & $0.18 \pm 0.01$ & \checkmark \\
ZTF J1905+3134 & 19:05:11.36 +31:34:32.3 & 17.2 & High & $20.70 \pm 0.02$ & -- & \checkmark \\
SDSS J1908+3940 & 19:08:17.07 +39:40:36.4 & 18.2 & High & $16.22 \pm 0.00$ & -- & \checkmark \\
HP Lib & 15:35:53.08 -14:13:12.2 & 18.4 & High & $13.60 \pm 0.00$ & $0.07 \pm 0.00$ & \checkmark \\
ASASSN-14cc & 21:39:48.24 -59:59:32.4 & 22.5(sh) & Outburst & $18.13 \pm 0.07$ & -- & $\times$ \\
PTF1 J1919+4815 & 19:19:05.19 +48:15:06.2 & 22.5 & Outburst & $19.75 \pm 0.05$ & -- & \checkmark \\
MGAB-V240 & 18:55:29.82 +32:30:17.8 & 22.8(sh) & Outburst & $18.69 \pm 0.03$ & -- & $\times$ \\
CXOGBS J1751-2940 & 17:51:07.60 -29:40:37.0 & 22.9 & High & $17.47 \pm 0.00$ & $0.07 \pm 0.01$ & $\times$ \\
TIC 378898110 & 12:03:38.70 -60:22:48.0 & 23.0? (sh?) & High & $14.28 \pm 0.00$ & -- & \checkmark \\
SDSS J1831+4202 & 18:31:31.63 +42:02:20.2 & 23.1(sh?) & High? & $17.55 \pm 0.03$ & -- & \checkmark \\
3XMM J0510-6703 & 05:10:34.60 -67:03:53.0 & 23.6 & Direct & $20.92 \pm 0.03$ & -- & \checkmark \\
CR Boo & 13:48:55.22 +07:57:35.8 & 24.5 & Outburst & $15.47 \pm 0.05$ & $0.06 \pm 0.00$ & \checkmark \\
KL Dra & 19:24:38.28 +59:41:46.7 & 25.0 & Outburst & $19.10 \pm 0.03$ & $0.09 \pm 0.00$ & \checkmark \\
ZTF J0729-0602 & 07:29:07.69 -06:02:46.6 & 25.9 & Outburst & $20.41 \pm 0.01$ & -- & $\times$ \\
V803 Cen & 13:23:44.54 -41:44:29.5 & 26.6 & Outburst & $15.73 \pm 0.11$ & $0.06 \pm 0.01$ & \checkmark \\
PTF1 J0719+4858 & 07:19:12.13 +48:58:34.0 & 26.8 & Outburst & $18.93 \pm 0.02$ & $0.15 \pm 0.01$ & \checkmark \\
WD J1040-4951 & 10:40:19.50 -49:51:29.7 & 27.5(sh?) & Outburst & $17.11 \pm 0.00$ & -- & $\times$ \\
ASASSN-15kf & 15:38:38.24 -30:35:49.7 & 27.7(sh) & Outburst & $19.40 \pm 0.01$ & -- & $\times$ \\
PTF1 J2219+3135 & 22:19:10.09 +31:35:23.1 & 27.7? (sh?) & Outburst & $19.31 \pm 0.08$ & -- & \checkmark \\
YZ LMi & 09:26:38.71 +36:24:02.4 & 28.3 & Outburst & $19.27 \pm 0.00$ & $0.04 \pm 0.00$ & \checkmark \\
CP Eri & 03:10:32.76 -09:45:05.3 & 28.4 & Outburst & $19.87 \pm 0.02$ & $0.05 \pm 0.01$ & \checkmark \\
ZTF J2228+4949 & 22:28:27.06 +49:49:16.5 & 28.6(sh?) & High & $19.24 \pm 0.00$ & -- & \checkmark \\
CRTS J0910-2008 & 09:10:17.45 -20:08:12.5 & 29.7(sh) & Outburst & $19.37 \pm 0.01$ & -- & \checkmark \\
Gaia16all & 06:27:20.53 -75:13:33.4 & 30.1(sh?) & Outburst & $20.55 \pm 0.02$ & -- & $\times$ \\
PTF1 J0943+1029 & 09:43:29.59 +10:29:57.6 & 30.4 & Outburst & $20.79 \pm 0.01$ & -- & \checkmark \\
ASASSN-19ct & 11:33:16.00 -37:10:20.0 & 31.2(sh) & Outburst & -- & -- & $\times$ \\
CRTS J0105+1903 & 01:05:50.10 +19:03:17.2 & 31.6? & Outburst & $19.72 \pm 0.01$ & -- & $\times$ \\
ZTF J0701+5023 & 07:01:15.77 +50:23:21.5 & 32.5 & Outburst & $20.50 \pm 0.01$ & -- & $\times$ \\
V406 Hya & 09:05:54.79 -05:36:08.6 & 33.8 & Outburst & $20.19 \pm 0.01$ & $0.04 \pm 0.01$ & \checkmark \\
PTF1 J0435+0029 & 04:35:17.73 +00:29:40.7 & 34.3 & Outburst & -- & -- & \checkmark \\
SDSS J1730+5545 & 17:30:47.59 +55:45:18.5 & 35.2 & Outburst & $20.05 \pm 0.01$ & -- & \checkmark \\
ZTF J0407-0007 & 04:07:49.30 -00:07:16.7 & 35.4 & Outburst & -- & $0.02 \pm 0.00$ & \checkmark \\
ASASSN-21hc & 16:05:25.23 -38:12:11.4 & 35.8? (sh?) & Outburst & $19.45 \pm 0.00$ & -- & \checkmark \\
2QZ J1427-01 & 14:27:01.70 -01:23:10.0 & 36.6(sh) & Outburst & $20.36 \pm 0.01$ & -- & $\times$ \\
NSV 1440 & 03:55:17.83 -82:26:11.5 & 36.6 & Outburst & $18.39 \pm 0.01$ & $0.04 \pm 0.00$ & \checkmark \\
ASASSN-21br & 16:18:10.48 -51:54:16.5 & 36.7(sh) & Outburst & $19.70 \pm 0.02$ & -- & \checkmark \\
SDSS J1240-0159 & 12:40:58.03 -01:59:19.2 & 37.4 & Outburst & $19.57 \pm 0.01$ & $0.04 \pm 0.01$ & \checkmark \\
ZTF J2252-0519 & 22:52:37.10 -05:19:17.4 & 37.4 & Outburst & $19.04 \pm 0.01$ & $0.03 \pm 0.01$ & \checkmark \\
SDSS J0129+3842 & 01:29:40.06 +38:42:10.5 & 37.6 & Outburst & $19.92 \pm 0.01$ & $0.05 \pm 0.02$ & \checkmark \\
SDSSJ1721+2733 & 17:21:02.48 +27:33:01.2 & 38.1 & Outburst & $20.25 \pm 0.01$ & -- & \checkmark \\
ASASSN-15oe & 00:11:59.39 -55:41:51.8 & 38.3(sh) & Outburst & $19.53 \pm 0.03$ & -- & $\times$ \\
ZTF J1838+0744 & 18:38:47.16 +07:44:46.2 & 39.6 & Outburst & $20.08 \pm 0.02$ & -- & $\times$ \\
ASASSN-14mv & 07:13:27.28 +20:55:53.4 & 40.9(sh) & Outburst & $18.02 \pm 0.01$ & $0.03 \pm 0.00$ & \checkmark \\
ASASSN-14ei & 02:55:33.39 -47:50:42.0 & 42.0? & Outburst & $16.46 \pm 0.02$ & -- & \checkmark \\
SDSS J1525+3600 & 15:25:09.58 +36:00:54.6 & 44.3 & Outburst & $19.78 \pm 0.01$ & -- & \checkmark \\
SDSS J0804+1616 & 08:04:49.49 +16:16:24.8 & 44.5 & Low & $18.35 \pm 0.01$ & -- & \checkmark \\
SDSS J1411+4812 & 14:11:18.31 +48:12:57.6 & 46.0 & Outburst & $19.55 \pm 0.00$ & -- & \checkmark \\
GP Com & 13:05:42.43 +18:01:04.0 & 46.5 & Low & -- & $0.02 \pm 0.00$ & \checkmark \\
CRTS J0450-0931 & 04:50:19.82 -09:31:12.8 & 47.3(sh) & Outburst & -- & -- & $\times$ \\
\qquad \vdots & \vdots & \vdots & \vdots & \vdots & \vdots & \vdots \\
\hline
\end{tabular}
\end{table*}

\begin{table*}
\caption{\textit{cont.---}List of confirmed AM\,CVn-type (without detected hydrogen) systems in the catalogue.
\label{tab:cat2}
}
\begin{tabular}{lcccccc}
\hline
Name & Coordinates & $P_{\rm orb}$ & Disc state & $G$ mag. & $q$ & Has \\
& & [min] &&& & spectrum \\
\hline
SDSS J0902+3819 & 09:02:21.36 +38:19:41.9 & 48.3 & Outburst & $20.12 \pm 0.01$ & $0.03 \pm 0.02$ & \checkmark \\
Gaia14aae & 16:11:33.97 +63:08:31.8 & 49.7 & Outburst & $18.29 \pm 0.03$ & $0.03 \pm 0.00$ & \checkmark \\
PNV J0624+0208 & 06:24:52.97 +02:08:20.7 & 50.4? (sh) & Outburst & $19.23 \pm 0.01$ & -- & $\times$ \\
ASASSN-17fp & 18:08:51.10 -73:04:04.2 & 51.0(sh) & Outburst & -- & -- & \checkmark \\
SDSS J1208+3550 & 12:08:41.96 +35:50:25.2 & 53.0 & Low & $18.83 \pm 0.00$ & -- & \checkmark \\
SDSS J0807+4852 & 08:07:10.33 +48:52:59.6 & 53.0(sh) & Low & $20.69 \pm 0.02$ & -- & \checkmark \\
ZTF J0220+2141 & 02:20:08.60 +21:41:55.8 & 53.4 & Low & $19.72 \pm 0.01$ & $0.02 \pm 0.01$ & \checkmark \\
SDSS J1642+1934 & 16:42:28.08 +19:34:10.1 & 54.2 & Low & $20.34 \pm 0.01$ & -- & \checkmark \\
ZTF J0003+1404 & 00:03:22.40 +14:04:59.0 & 55.5 & Outburst & $20.00 \pm 0.02$ & $0.02 \pm 0.00$ & \checkmark \\
SDSS J1552+3201 & 15:52:52.48 +32:01:50.9 & 56.3 & Low & $20.14 \pm 0.02$ & -- & \checkmark \\
SRGe J0453+6224 & 04:53:59.90 +62:24:44.0 & 58.1 & Low & $18.58 \pm 0.01$ & $0.05 \pm 0.02$ & \checkmark \\
ASASSN-21au & 14:23:52.82 +78:30:13.4 & 58.4(sh)? & Outburst & $20.92 \pm 0.02$ & -- & \checkmark \\
SDSS J1137+4054 & 11:37:32.32 +40:54:58.3 & 59.6? & Outburst & $19.12 \pm 0.00$ & -- & \checkmark \\
ZTF J1637+4917 & 16:37:43.55 +49:17:40.9 & 61.5 & Low & $19.35 \pm 0.00$ & $0.03 \pm 0.01$ & \checkmark \\
V396 Hya & 13:12:46.93 -23:21:31.3 & 65.1 & Low & -- & $0.01 \pm 0.00$ & \checkmark \\
SDSS J1319+5915 & 13:19:54.47 +59:15:14.8 & 65.6 & Low & $19.11 \pm 0.00$ & -- & \checkmark \\
SDSS J1505+0659 & 15:05:51.58 +06:59:48.7 & 68.4 & Low & $19.08 \pm 0.00$ & $0.01 \pm 0.00$ & \checkmark \\
1eRASS J1013-2028 & 10:13:28.7 -20:28:48 & -- & Low & $18.27 \pm 0.00$ & -- & \checkmark \\
ASASSN-14fv & 23:29:55.13 +44:56:14.4 & -- & Outburst & $19.85 \pm 0.10$ & -- & \checkmark \\
CRTS J0744+3254 & 07:44:19.70 +32:54:48.0 & -- & Outburst & $19.83 \pm 0.02$ & -- & \checkmark \\
CRTS J0844-0128 & 08:44:13.60 -01:28:07.0 & -- & Outburst & $20.37 \pm 0.02$ & -- & \checkmark \\
MGAB-V270 & 21:28:22.20 +63:25:57.2 & -- & Outburst & $19.99 \pm 0.01$ & -- & \checkmark \\
MOA 2010-BLG-087 & 18:08:34.85 -26:29:22.8 & -- & Outburst & $19.50 \pm 0.02$ & -- & $\times$ \\
PTF1 J0857+0729 & 08:57:24.27 +07:29:46.7 & -- & Outburst & -- & -- & \checkmark \\
PTF1 J1523+1845 & 15:23:10.71 +18:45:58.2 & -- & Outburst & -- & -- & \checkmark \\
PTF1 J1632+3511 & 16:32:39.39 +35:11:07.3 & -- & Outburst & -- & -- & \checkmark \\
SDSS J0129-5808 & 01:29:05.53 -58:08:41.2 & -- & Low & $19.48 \pm 0.00$ & -- & \checkmark \\
SDSS J0903-0133 & 09:03:44.25 -01:33:26.1 & -- & Low & $19.53 \pm 0.01$ & -- & \checkmark \\
SDSS J0953-0427 & 09:53:12.66 -04:27:32.0 & -- & Outburst & -- & -- & \checkmark \\
SDSS J1043+5632 & 10:43:25.08 +56:32:58.1 & -- & Outburst & $20.47 \pm 0.01$ & -- & \checkmark \\
SDSS J1917-0951 & 19:17:05.21 -09:51:44.8 & -- & Outburst & $18.77 \pm 0.01$ & -- & \checkmark \\
SDSS J2047+0008 & 20:47:39.40 +00:08:40.3 & -- & Low & -- & -- & \checkmark \\
ZTF18aavetqn & 19:18:42.00 +44:49:12.3 & -- & Outburst & $19.54 \pm 0.01$ & -- & \checkmark \\
ZTF18acbdbhx & 21:08:20.62 -13:49:09.4 & -- & Outburst & $19.72 \pm 0.02$ & -- & \checkmark \\
ZTF18acnnabo & 08:20:47.63 +68:04:23.9 & -- & Outburst & $20.19 \pm 0.01$ & -- & \checkmark \\
ZTF18acujsfl & 04:49:30.07 +02:51:53.7 & -- & Outburst & $20.30 \pm 0.01$ & -- & \checkmark \\
ZTF19aaktdwc & 13:29:18.49 -12:16:22.6 & -- & Outburst & $20.38 \pm 0.01$ & -- & \checkmark \\
ZTF19abzzuin & 08:44:19.72 +06:39:50.2 & -- & Outburst & $21.16 \pm 0.03$ & -- & \checkmark \\
eFEDS J0847+0119 & 08:47:08.30 +01:19:39.5 & -- & Low & -- & -- & \checkmark \\
\hline
\end{tabular}
\end{table*}

\begin{table*}
\caption{List of non-AM\,CVn confirmed UCBs in the catalogue.
\label{tab:cat3}
}
\begin{tabular}{lcccccc}
\hline
Name & Coordinates & $P_{\rm orb}$ & Disc state & $G$ mag. & $q$ & Has \\
& & [min] &&& & spectrum \\
\hline
\textit{He~CVs} \\
MASTER OT J2348+2502 & 23:48:43.23 +25:02:50.4 & 49.0(sh) & Outburst & $20.08 \pm 0.01$ & -- & \checkmark \\
ZTF J1813+4251 & 18:13:11.13 +42:51:50.4 & 51.2 & Outburst & $18.92 \pm 0.01$ & $0.21 \pm 0.01$ & \checkmark \\
KSP-OT-201701a & 06:39:23.2 -26:37:18.8 & 51.9 & Outburst & $18.59 \pm 0.08$ & $0.40 \pm 0.24$ & \checkmark \\
CRTS J1028-0819 & 10:28:43 -08:19:27 & 53.4 & Outburst & $19.32 \pm 0.01$ & $0.25 \pm 0.06$ & \checkmark \\
CRTS J1111+5712 & 11:11:26.9 +57:12:39 & 55.3 & Outburst & $18.85 \pm 0.02$ & $0.09 \pm 0.01$ & \checkmark \\
ZTF J2116+2446 & 21:16:04.73 +24:46:20.5 & 56.2 & Outburst & $20.64 \pm 0.02$ & -- & $\times$ \\
BOKS45906 & 19:40:16.2 +46:32:47.9 & 56.4 & Outburst & $19.97 \pm 0.02$ & -- & \checkmark \\
KSP-OT-201712a & 07:24:07.70 -26:19:54.5 & 58.8 & Outburst & $16.98 \pm 0.02$ & -- & \checkmark \\
V485 Cen & 12:57:23.3 -33:12:07 & 59.0 & Outburst & $17.91 \pm 0.01$ & $0.38 \pm 0.13$ & \checkmark \\
ASASSN-22ak & 22:56:08.86 -68:36:10.2 & 61.7(sh) & Outburst & $20.43 \pm 0.01$ & -- & $\times$ \\
CRTS J2333-1557 & 23:33:13 -15:57:44 & 62.0 & Outburst & $19.85 \pm 0.02$ & -- & $\times$ \\
EI Psc & 23:29:54.3 +06:28:11 & 64.2 & Outburst & $15.88 \pm 0.01$ & $0.18 \pm 0.01$ & \checkmark \\
CRTS 1740+4147 & 17:40:33.46 +41:47:55.98 & 64.8 & Outburst & $19.60 \pm 0.01$ & $0.05 \pm 0.01$ & $\times$ \\
CRTS J1122-1110 & 11:22:53.3 -11:10:37 & 65.2 & Outburst & $20.30 \pm 0.01$ & $0.02 \pm 0.01$ & \checkmark \\
V418 Ser & 15:14:53.64 +02:09:34.54 & 65.9 & Outburst & $20.15 \pm 0.01$ & -- & \checkmark \\
\\
\textit{sdB donor stars} \\
ZTF J2130+4420 & 21:30:56.71 +44:20:46.5 & 39.3 & ?? & $15.44 \pm 0.01$ & $0.62 \pm 0.04$ & \checkmark \\
ZTF J2055+4651 & 20:55:15.98 +46:51:06.5 & 56.3 & ?? & $17.61 \pm 0.01$ & $0.60 \pm 0.07$ & \checkmark \\
\\
\textit{Brown dwarf donor stars} \\
SDSS J1507+5230 & 15:07:22.33 +52:30:39.8 & 66.6 & Outburst & $18.15 \pm 0.01$ & $0.06 \pm 0.00$ & \checkmark \\
WD J1540-3929 & 15:40:08.28 -39:29:17.6 & 68.4 & ?? & $17.33 \pm 0.00$ & -- & \checkmark \\
\\
\textit{Others} \\
ZTF J0127+5258 & 01:27:47.62 +52:58:13.0 & 13.7 & ?? & $19.81 \pm 0.01$ & -- & \checkmark \\
ATLAS J1138-5139 & 11:38:10.91 -51:39:49.2 & 27.7 & ?? & $16.89 \pm 0.01$ & $0.23 \pm 0.03$ & \checkmark \\
\\
\textit{Unclear} \\
ASASSN-19rg & 13:25:58.13 -14:52:26.3 & 43.9? (sh) & Outburst & $20.38 \pm 0.01$ & -- & $\times$ \\
ASASSN-18rg & 21:17:42.19 -02:22:27.6 & 46.0? (sh?) & Outburst & $20.51 \pm 0.01$ & -- & $\times$ \\
V2276 Sgr & 20:26:22.19 -43:40:31.7 & 28.3? & Outburst & $19.96 \pm 0.01$ & -- & \checkmark \\
ASASSN-16hb & 14:43:25.56 +79:11:59.6 & 35.6? & Outburst & $19.26 \pm 0.06$ & -- & $\times$ \\
TCP J0427-2528 & 04:27:21.95 -25:28:03.8 & 43.6? (sh) & Outburst & $21.07 \pm 0.03$ & -- & $\times$ \\
TCP J0722+6220 & 07:22:26.83 +62:20:54.8 & 46.9? (sh) & Outburst & $20.01 \pm 0.01$ & -- & $\times$ \\
\hline
\end{tabular}
\end{table*}

\begin{table*}
\caption{List of candidate UCBs in the catalogue.
\label{tab:cat4}
}
\begin{tabular}{lcccccc}
\hline
Name & Coordinates & $P_{\rm orb}$ & Disc state & $G$ mag. & $q$ & Has \\
& & [min] &&& & spectrum \\
\hline
N1851-FUV1 & 05:14:06.76 -40:02:47.6 & 18.1(sh?) & ?? & -- & -- & $\times$ \\
ASASSN-18dh & 11:48:53.65 -03:11:54.6 & -- & Outburst & -- & -- & $\times$ \\
ASASSN-18vi & 18:31:13.13 +41:40:29.1 & -- & Outburst & -- & -- & $\times$ \\
ASASSN-19hs & 06:19:52.21 +24:20:58.6 & -- & Outburst & $20.36 \pm 0.03$ & -- & $\times$ \\
ASASSN-20eq & 17:35:00.51 +25:36:55.4 & -- & Outburst & -- & -- & $\times$ \\
ASASSN-20gx & 23:49:30.14 +22:01:29.6 & -- & Outburst & $20.19 \pm 0.01$ & -- & $\times$ \\
ASASSN-20jt & 23:02:35.94 +53:28:21.3 & -- & Outburst & $19.96 \pm 0.03$ & -- & $\times$ \\
ASASSN-20ke & 18:42:17.72 +16:55:00.7 & -- & Outburst & $20.48 \pm 0.03$ & -- & $\times$ \\
ASASSN-20la & 01:38:51.92 +46:34:51.3 & -- & Outburst & -- & -- & $\times$ \\
ASASSN-20lr & 04:22:20.03 +50:07:12.7 & -- & Outburst & $19.66 \pm 0.01$ & -- & $\times$ \\
ASASSN-21fhx & 21:04:11.01 -01:02:02.3 & -- & Outburst & $19.40 \pm 0.00$ & -- & $\times$ \\
ASASSN-21uh & 19:34:30.71 +47:39:06.9 & -- & Outburst & -- & -- & $\times$ \\
CRTS J0808+3550 & 08:08:53.7 +35:50:53 & -- & Outburst & $19.98 \pm 0.03$ & -- & \checkmark \\
CRTS J1647+4338 & 16:47:48.0 +43:38:45 & -- & Outburst & $20.02 \pm 0.03$ & -- & \checkmark \\
CRTSJ0850-2358 & 08:50:24.20 -23:58:11.0 & -- & Outburst & $20.18 \pm 0.01$ & -- & $\times$ \\
CRTSJ2052-0614 & 20:52:52.30 -06:14:41.0 & -- & Outburst & $20.77 \pm 0.01$ & -- & $\times$ \\
Gaia21fgn & 09:47:08.86 +23:31:51.6 & -- & Outburst & $20.21 \pm 0.01$ & -- & $\times$ \\
MASTER OT J0558+3915 & 05:58:45.55 +39:15:33.4 & -- & Outburst & $18.70 \pm 0.01$ & -- & $\times$ \\
MGAB-V298 & 06:23:48.18 +40:49:42.6 & -- & Outburst & $19.75 \pm 0.11$ & -- & $\times$ \\
MGAB-V300 & 19:49:47.59 +03:13:05.2 & -- & Outburst & $19.01 \pm 0.01$ & -- & $\times$ \\
MGAB-V500 & 00:06:27.21 +49:10:29.5 & -- & Outburst & $18.50 \pm 0.04$ & -- & $\times$ \\
PS1-3PI J0633-2305 & 06:33:34.87 -23:05:46.4 & -- & Outburst & $18.76 \pm 0.01$ & -- & $\times$ \\
PS16bqk & 18:40:43.89 +27:01:25.0 & -- & Outburst & $18.91 \pm 0.32$ & -- & $\times$ \\
TCP J0050+5351 & 00:50:56.44 +53:51:52.4 & -- & Outburst & $18.31 \pm 0.00$ & -- & $\times$ \\
TCP J0219+3735 & 02:19:42.36 +37:35:54.0 & -- & Outburst & -- & -- & $\times$ \\
TCP J0609+3046 & 06:09:37.84 +30:46:51.9 & -- & Outburst & $20.18 \pm 0.01$ & -- & $\times$ \\
TCPJ0028+4232 & 00:28:04.58 +42:32:54.4 & -- & Outburst & -- & -- & $\times$ \\
W62-UV & 00:23:59.34 -72:04:48.3 & -- & ?? & -- & -- & $\times$ \\
ZTF22abnmsdh & 23:30:47.87 +25:25:57.3 & -- & Outburst & -- & -- & $\times$ \\
\hline
\end{tabular}
\end{table*}

\section{References}

Table~\ref{tab:refs} contains the reference keys used in the catalogue and links them to the full reference listed in the reference section.

\begin{table}
\caption{References used in the catalogue.
\label{tab:refs}
}
\begin{tabular}{lc}
\hline
Abbreviation & Reference\\
\hline
Abbott1992 & \citet{Abbott1992}\\
Anderson2005 & \citet{Anderson2005}\\
Anderson2008 & \citet{Anderson2008}\\
Armstrong2012 & \citet{Armstrong2012}\\
Augusteijn1993 & \citet{Augusteijn1993}\\
Augusteijn1996 & \citet{Augusteijn1996}\\
Aungwerojwit2025 & \citet{Aungwerojwit2025}\\
Aydi2021 & \citet{Aydi2021}\\
Bakowska2021 & \citet{Bakowska2021}\\
Barros2007 & \citet{Barros2007}\\
Boneva2022 & \citet{Boneva2022}\\
Breedt2012 & \citet{Breedt2012}\\
Breedt2014 & \citet{Breedt2014}\\
Burdge2020 & \citet{Burdge2020}\\
Burdge2022 & \citet{Burdge2022}\\
Burdge2023 & \citet{Burdge2023}\\
Campbell2015 & \citet{Campbell2015}\\
Carter2013 & \citet{Carter2013}\\
Carter2014a & \citet{Carter2014a}\\
Carter2014b & \citet{Carter2014b}\\
Chakraborty2024 & \citet{Chakraborty2024}\\
Chickles2024 & \citet{Chickles2024}\\
Chochol2015 & \citet{Chochol2015}\\
Copperwheat2011a & \citet{Copperwheat2011a}\\
Copperwheat2011b & \citet{Copperwheat2011b}\\
Cropper1998 & \citet{Cropper1998}\\
Denisenko2014 & \citet{Denisenko2014}\\
Deshmukh2023 & \citet{Deshmukh2023}\\
Espaillat2005 & \citet{Espaillat2005}\\
Fontaine2011 & \citet{Fontaine2011}\\
Garnavich2014 & \citet{Garnavich2014}\\
Green2018a & \citet{Green2018a}\\
Green2018b & \citet{Green2018b}\\
Green2019 & \citet{Green2019}\\
Green2020 & \citet{Green2020}\\
Green2024 & \citet{Green2024}\\
Haberl2017 & \citet{Haberl2017}\\
Han2021 & \citet{Han2021}\\
Hardy2017 & \citet{Hardy2017}\\
Harvey1998 & \citet{Harvey1998}\\
Imada2018 & \citet{Imada2018}\\
Inight2023a & \citet{Inight2023a}\\
Inight2023b & \citet{Inight2023b}\\
Isogai2016 & \citet{Isogai2016}\\
Isogai2019 & \citet{Isogai2019}\\
Isogai2021 & \citet{Isogai2021}\\
Israel1999 & \citet{Israel1999}\\
Israel2002 & \citet{Israel2002}\\
Jha1998 & \citet{Jha1998}\\
Kato2009 & \citet{Kato2009}\\
Kato2014 & \citet{Kato2014}\\
Kato2015 & \citet{Kato2015}\\
Kato2021 & \citet{Kato2021}\\
Kato2023a & \citet{Kato2023a}\\
Kato2023b & \citet{Kato2023b}\\
Kato2023c & \citet{Kato2023c}\\
Kepler2016 & \citet{Kepler2016}\\
Khalil2024 & \citet{Khalil2024}\\
Kong2018 & \citet{Kong2018}\\
\qquad \vdots & \vdots \\
\hline
\end{tabular}
\end{table}

\begin{table}
\caption{References used in the catalogue, \textit{cont.}}
\begin{tabular}{lc}
\hline
Abbreviation & Reference\\
\hline
Kosakowski2024 & \citet{Kosakowski2024}\\
Kupfer2013 & \citet{Kupfer2013}\\
Kupfer2015 & \citet{Kupfer2015}\\
Kupfer2016 & \citet{Kupfer2016}\\
Kupfer2019 & \citet{Kupfer2019}\\
Kupfer2020a & \citet{Kupfer2020a}\\
Kupfer2020b & \citet{Kupfer2020b}\\
Lee2022 & \citet{Lee2022}\\
Lee2024 & \citet{Lee2024}\\
Levitan2011 & \citet{Levitan2011}\\
Levitan2013 & \citet{Levitan2013}\\
Levitan2014 & \citet{Levitan2014}\\
Levitan2015 & \citet{Levitan2015}\\
Littlefair2007 & \citet{Littlefair2007}\\
Littlefield2013 & \citet{Littlefield2013}\\
Maitra2024 & \citet{Maitra2024}\\
Marcano2021 & \citet{Marcano2021}\\
Marsh1999 & \citet{Marsh1999}\\
Marsh2017 & \citet{Marsh2017}\\
Mason2003 & \citet{Mason2003}\\
Motch1996 & \citet{Motch1996}\\
Munday2023 & \citet{Munday2023}\\
Munday2024 & \citet{Munday2024}\\
Nather1981 & \citet{Nather1981}\\
Nather1984 & \citet{Nather1984}\\
Nelemans2001a & \citet{Nelemans2001a}\\
Nesci2013 & \citet{Nesci2013}\\
O'Donoghue1987 & \citet{O'Donoghue1987}\\
O'Donoghue1994 & \citet{O'Donoghue1994}\\
Painter2024 & \citet{Painter2024}\\
Patterson1997 & \citet{Patterson1997}\\
Patterson2000 & \citet{Patterson2000}\\
Patterson2002 & \citet{Patterson2002}\\
Prieto2014 & \citet{Prieto2014}\\
Provencal1991 & \citet{Provencal1991}\\
Provencal1994 & \citet{Provencal1994}\\
Ramsay2000 & \citet{Ramsay2000}\\
Ramsay2002 & \citet{Ramsay2002}\\
Ramsay2010 & \citet{Ramsay2010}\\
Ramsay2014 & \citet{Ramsay2014}\\
Ramsay2018 & \citet{Ramsay2018}\\
Ramsay2018b & \citet{Ramsay2018b}\\
Rau2010 & \citet{Rau2010}\\
RiveraSandoval2018 & \citet{RiveraSandoval2018}\\
RiveraSandoval2019 & \citet{RiveraSandoval2019}\\
RiveraSandoval2020 & \citet{RiveraSandoval2020}\\
RiveraSandoval2021 & \citet{RiveraSandoval2021}\\
RiveraSandoval2022 & \citet{RiveraSandoval2022}\\
Rodriguez2023 & \citet{Rodriguez2023}\\
Rodriguez2024 & \citet{Rodriguez2024}\\
Roelofs2005 & \citet{Roelofs2005}\\
Roelofs2006a & \citet{Roelofs2006a}\\
Roelofs2006b & \citet{Roelofs2006b}\\
Roelofs2007b & \citet{Roelofs2007b}\\
Roelofs2009 & \citet{Roelofs2009}\\
Roelofs2010 & \citet{Roelofs2010}\\
RoelofsPhD & \citet{RoelofsPhD}\\
Ruiz2001 & \citet{Ruiz2001}\\
Schwope2024a & \citet{Schwope2024a}\\
\qquad \vdots & \vdots \\
\hline
\end{tabular}
\end{table}

\begin{table}
\caption{References used in the catalogue, \textit{cont.}}
\begin{tabular}{lc}
\hline
Abbreviation & Reference\\
\hline
Skillman1999 & \citet{Skillman1999}\\
Smak1967 & \citet{Smak1967}\\
Smak2023 & \citet{Smak2023}\\
Soraisam2023 & \citet{Soraisam2023}\\
Steeghs2006 & \citet{Steeghs2006}\\
Steen2024 & \citet{Steen2024}\\
Strohmayer2004 & \citet{Strohmayer2004}\\
Strohmayer2021 & \citet{Strohmayer2021}\\
Szkody2005 & \citet{Szkody2005}\\
Szkody2020 & \citet{Szkody2020}\\
Wagner2014 & \citet{Wagner2014}\\
Warner1972 & \citet{Warner1972}\\
Warner2002 & \citet{Warner2002}\\
Wong2021b & \citet{Wong2021b}\\
Woudt2003 & \citet{Woudt2003}\\
Woudt2004 & \citet{Woudt2004}\\
Woudt2005 & \citet{Woudt2005}\\
Woudt2011 & \citet{Woudt2011}\\
Woudt2013 & \citet{Woudt2013}\\
Zurek2016 & \citet{Zurek2016}\\
deMiguel2018 & \citet{deMiguel2018}\\
vanRoestel2021 & \citet{vanRoestel2021}\\
vanRoestel2022 & \citet{vanRoestel2022}\\
\hline
\end{tabular}
\end{table}

\end{appendix}

\end{document}